\documentclass[aps,10pt,twocolumn,superscriptaddress]{revtex4-1}

\usepackage{amsmath, bm, amsfonts, amssymb}     
\usepackage{graphicx}                       
\usepackage{xfrac}                          
\usepackage{grffile}                        
\usepackage{color,colortbl}
\usepackage{xcolor}
\usepackage{ wasysym }
\usepackage{placeins}

\usepackage[colorlinks, citecolor=blue, urlcolor=blue, linkcolor=blue]{hyperref}

\begin{document}
	
	\title{Confinement-induced demixing and crystallization}
	
	\author{Gerhard Jung}
		\email{Corresponding author: gerhard.jung@uibk.ac.at}
	\affiliation{Institut f\"ur Theoretische Physik, Universit\"at Innsbruck, Technikerstra{{\ss}}e 21A, A-6020 Innsbruck, Austria}
	\author{Charlotte F. Petersen}
	\affiliation{Institut f\"ur Theoretische Physik, Universit\"at Innsbruck, Technikerstra{{\ss}}e 21A, A-6020 Innsbruck, Austria}
	
	\begin{abstract}
		
		We simulate a strongly size-disperse hard-sphere fluid confined between two parallel, hard walls. We find that confinement induces crystallization into $ n $-layered hexagonal lattices and a novel honeycomb-shaped structure, facilitated by fractionation. The onset of freezing prevents the formation of a stable glass phase and occurs at much smaller packing fraction than in bulk.  Varying the wall separation triggers solid-to-solid transitions and a systematic change of the size-distribution of crystalline particles, which we rationalize using a semi-quantitative theory. We show that the crystallization can be exploited in a wedge geometry to demix particles of different sizes. 
		
	\end{abstract}
	
	\maketitle
	
	\section{Introduction}
	
	Confinement occurs naturally in many physical, chemical and biological systems like nanotubes, porous rocks or crowded living cells.  Even for the simplest case of monodisperse hard spheres confined between parallel, hard walls one observes spatially inhomogeneous density profiles and diffusivities \cite{theo:Kjellander1988,sim:Mittal2008}, anisotropic structure factors \cite{ theo:Gotzelmann1997,sim:Mandal2017a},  multiple-reentrant glass transitions \cite{theo:Lang2010,sim:Mandal2014} and solid-to-solid transitions between different crystalline phases \cite{theo:Schmidt1996,theo:Schmidt1997,sim:Fortini_2006,sim:Oguz2012}. Confinement has also been reported to have a strong impact on the structural relaxation of supercooled liquids \cite{sim:Scheidler2002, sim:Mandal2014,Roberts2020_dyn}, since it restricts the range of accessible length scales, which becomes increasingly important as the glass transition is approached \cite{sim:Scheidler2002}.  The effect of confinement on the properties of simple liquids or colloids has therefore been carefully studied with experiments \cite{exp:Nugent2007,exp:Sarangapani2011,exp:Reinmuller2012,exp:nygard2012, exp:nygard2013,exp:Hunter2014,exp:Williams2015, exp:nygard2016_2, exp:Zhang2016, exp:Kienle2016, exp:Lippmann2019}, theory \cite{theo:Kjellander1988,theo:Schmidt1996,theo:Schmidt1997, theo:Gotzelmann1997,theo:Rosenfeld1997,theo:Roth2005, theo:Oguz2009}  and simulations \cite{sim:Fehr1995,sim:Rodriguez1996,sim:Scheidler2002,sim:Fortini_2006,sim:Mittal2007_2,sim:Mittal2008,sim:Goel2009,sim:Krishnan2012,sim:Oguz2012,sim:Ingebrigtsen2013,sim:Saw2016,Roberts2020_dyn}. The results indicate that the rich phenomenology of confined hard spheres indeed accurately applies to liquids and colloids in nanoscopic confinement. It has also been highlighted that these confinement-induced structural properties can strongly depend on the specifics of the system, such as the roughness of the walls \cite{sim:Scheidler2004} and the interaction between the particles \cite{PhysRevLett.109.028301}. 

	An important aspect has, however, not yet received much attention, which is the effect of particle size-dispersity on the above phenomenology. In laboratory experiments on colloids a size-dispersity occurs naturally from synthesis. Additionally, for the study of structural relaxation in supercooled liquids and glasses a size-dispersity must be introduced to prevent crystallization even at very small packing fractions \cite{exp:Nugent2007, sim:Mandal2014,Roberts2020_dyn}. Despite their popularity as model glass formers recent observations in bulk have revealed that  Gaussian-distributed hard spheres \cite{polycrist:Fasolo2003, polycrist:Zaccarelli2009, polycrist:Campo2016, polycrist:Bommineni2019} as well as the often used Kob-Anderson model \cite{kacrist:Hitchcock1999,kacrist:Pedersen2018} form crystals already in the supercooled regime. This crystallization is induced by a process called fractionation which describes the separation of a homogeneous fluid into different liquid or crystalline fractions with very different particle-size distributions. Since walls or other boundaries lead to heterogeneous crystallization which is known to strongly effect and often enhance nucleation processes \cite{hetcry:Sandomirski2011,hetcry:Espinosa2019} it should be expected that confinement could also have an impact on crystallization and fractionation. Understanding the equilibrium structural properties of confined size-disperse particles thus remains a critical outstanding problem and is essential for future studies of simple glass formers and colloids in confinement.
	
		In this work we study the crystallization of size-disperse hard spheres in a slit geometry. We use two complementary approaches: First, we perform exact event-driven molecular dynamics (EDMD) simulations with enhanced sampling techniques to reveal a complex phase behaviour, displaying amongst others a new phase which is shaped like a honeycomb lattice. Significantly, we find formation of stable crystals at much smaller packing fractions than reported for bulk systems, and show that the crystallization is enabled by a confinement-controlled fractionation. These surprising results show additionally that the multiple-reentrant glass transition reported in Ref.~\cite{sim:Mandal2014} is only metastable. Second, we rationalize our findings with a semi-quantitative theory allowing us to provide a deep insight into the mechanisms that drive the confinement-induced crystallization.  Most importantly, our study reveals a very general technique for the demixing of size-disperse particles with excluded volume interactions.
		
		\section{Methods}
		
		 The investigated system consists of two parallel, hard walls separated by a distance $ H $ in the $ z $-direction and periodic boundary conditions in the other two dimensions. The slab is filled with size-disperse hard spheres at packing fraction $ \phi $. In the spirit of recent simulations in bulk \cite{polycrist:Bommineni2019} we employ a Gaussian distribution with average particle diameter $ \sigma $ and dispersity $ \delta = 0.15 $. Here, $ \sigma $ sets the length scale of the system and the dispersity $ \delta $ is defined as the ratio of the standard deviation of the particle diameter to the average particle diameter. 
		
		\subsection{Event-driven molecular dynamics simulations}
		
The system is simulated using a hybrid scheme of event-driven molecular dynamics simulations (EDMD) in the \emph{NVT} ensemble \cite{edmd:alder1957,edmd:RAPAPORT1980,edmd:bannerman2011} and the recently proposed swap Monte Carlo algorithm (SWAP) \cite{swap:GAZZILLO1989,swap:ninarello2017,swap:Berthier2019}. SWAP introduces ``swap events'' that are triggered every $ \Delta t = 0.5\,\sigma v_\text{th}^{-1} $ time steps. Each swap event consists of $ N_s = 50,000 $ Monte Carlo moves in which two particles are randomly selected and their diameters are exchanged. The move is accepted if the particles exhibit no overlap with other particles after the swap. This corresponds to the typical Monte Carlo acceptance criterium for hard spheres and is consistent with detailed balance. The hybrid scheme is implemented in the EDMD code DynamO \cite{edmd:bannerman2011}. Its usage is essential to accelerate the dynamics and systematically ensure that the equilibrium state of the system is attained. We observe typical acceptance ratios of $ \Gamma= 0.05 $, showing that a significant proportion of the particles is swapped in each swap event. The slab consists of $ N= 13068$ particles for any wall separation $ 2.0 \leq H/\sigma \leq 3.0 $ resulting in a minimum longitudinal box size of $ L_x=L_y=67.15\,\sigma $, which is large enough to eliminate any finite size effects. The system is equilibrated for $ N_\text{e}= 3\cdot 10^{10} $ events, which roughly corresponds to a total equilibration time of $ t_\text{eq} = 7.0\cdot10^4\,\sigma v_\text{th}^{-1} $. 
		
		To determine the size-distribution of particles in the solid phase we define the rotationally invariant bond-order parameter $ d_6(m,n) = \phi_{6,m}\phi_{6,n} $.  Here, we introduced the local bond-orientational parameter,
		$ \phi_{6,m} = N_m^{-1} \sum_{m'=1}^{N_m} e^{6 \textrm{i} \theta_{mm'}},  $
		as the relative orientation of particle $ m $ with its $ N_m $ neighbors \cite{bo:Nelson1979, bo:Chakrabarti1998,bo:Pusey2009}. Neighbors are all particles $ m' $ within a cut-off distance $ r_\parallel < 1.4\,\sigma $ and $ r_\perp < 0.1\,\sigma $, with $ \perp $ denoting the direction perpendicular to the walls. Using these invariant measures we define a particle $ m $ as crystalline if it has at least four neighbours $ n $ for which the rotationally invariant bond-order parameter	$ d_6(m,n) > 0.7. $
		Above this threshold, the local environment of particle $ m $ is strongly correlated with the one of neighbor $ n $ which indicates a solid-like structure. To identify the big particles in the honeycomb phase $ 3\astrosun $ we set $ r_\parallel < 1.8\,\sigma $ and require only three crystalline neighbours.  We report the packing fraction $ \phi_c $ at which the onset of crystallization is observed, as well as the average particle diameter $ \bar{a} $ and dispersity $ \delta_c $ of particles in the solid phase \cite{bo:Nelson1979, bo:Chakrabarti1998,bo:Pusey2009}.\\
		
		\subsection{Theoretical model}
		\label{sec:theory}
		
		 The theoretical approach generalizes the technique proposed for monodisperse particles \cite{theo:Schmidt1997} by combining fundamental measure theory for size-disperse hard spheres in a slit geometry \cite{fmt:Rosenfeld1989,fmt:Roth2010,fmt:chemical2018} with cell theory \cite{cell:kirkwood1950,barker1963lattice,cell:Bonissent1984}.
		 
		 \subsubsection{Fundamental measure theory (FMT) }
		
		FMT is based on density functional theory and minimizes the functional for the grand free energy leading to the equation \cite{fmt:Rosenfeld1989,fmt:Roth2010},
		\begin{equation}\label{eq:fmt}
		\ln \lambda_i^3 n_i(z) = \beta \mu_i - \beta \frac{\delta \mathcal{F}^\text{ex}[n_i]}{\delta n_i(z)} - \beta V_i(z).
		\end{equation}
		Here, $ n_i(z) $ is the local number density, $ \mu_i $ the chemical potential \cite{fmt:chemical2018} and $ V_i(z) $ the external wall potential of component $ i $. Each component has a different diameter $ a_i = a_\text{min} + (i+0.5) \Delta a  $, which allows us to emulate a size-dispersity by creating a mixture of $ m $ different components, similar to Ref.~\cite{sim:Mandal2014}. We use $ m = 25 $ with $ \Delta a = 0.024\,\sigma $ and $ a_\text{min} = 0.7\,\sigma $. We also ensured that the density profiles and the resulting free energies only marginally change ($ \Delta \mathcal{F} < 0.02\,k_BT $) upon an increase of $ m $. For the excess free energy functional of a hard-sphere mixture $ \mathcal{F}^\text{ex} $ we use the White Bear Version II \cite{fmt:Roth2010}. To determine the density profile $ n_i(z) $, Eq.~(\ref{eq:fmt}) is solved self-consistently in an iterative procedure \cite{fmt:Roth2010}. Different from Ref.~\cite{sim:Mandal2014} we ensure in our algorithm that the chemical potential $ \mu_i $ is adapted in each iteration such that the resulting slit-averaged concentrations of every component $ \bar{n}_i(a_i) = \int \text{d}z\,n_i(z) $ precisely match a Gaussian profile with mean $ \sigma $ and variance $ \delta $. The free energy per particle of component $ i $ can then be evaluated from its ideal and excess contributions,
		\begin{equation}\label{eq:fe}
		\mathcal{F}_{\text{lq},i} = \frac{1}{\beta H n_i} \int \text{d}z n_i(z)(\ln \lambda_i^3 n_i(z) -1) + \frac{\mathcal{F}^\text{ex}}{Hn_i}.
		\end{equation}
		The constant $ \lambda_i $ is set to $ \lambda_i^3 = \sigma^4 \Delta a^{-1}  $, such that $ \rho_i(z) =  \lambda_i^3 n_i(z) = \sigma^4 \Delta a^{-1} n_i(z) $ is the density distribution of the disperse mixture in the spirit of Ref.~\cite{polycrist:Fasolo2003}. We carefully checked this expression and ensured that it is independent of $ a_\text{min} $ and $ \Delta a $ and indeed gives the correct ideal gas contributions (e.g. the expected values for the mixing free energy). 
		
			\begin{figure}
		\includegraphics[scale=0.9]{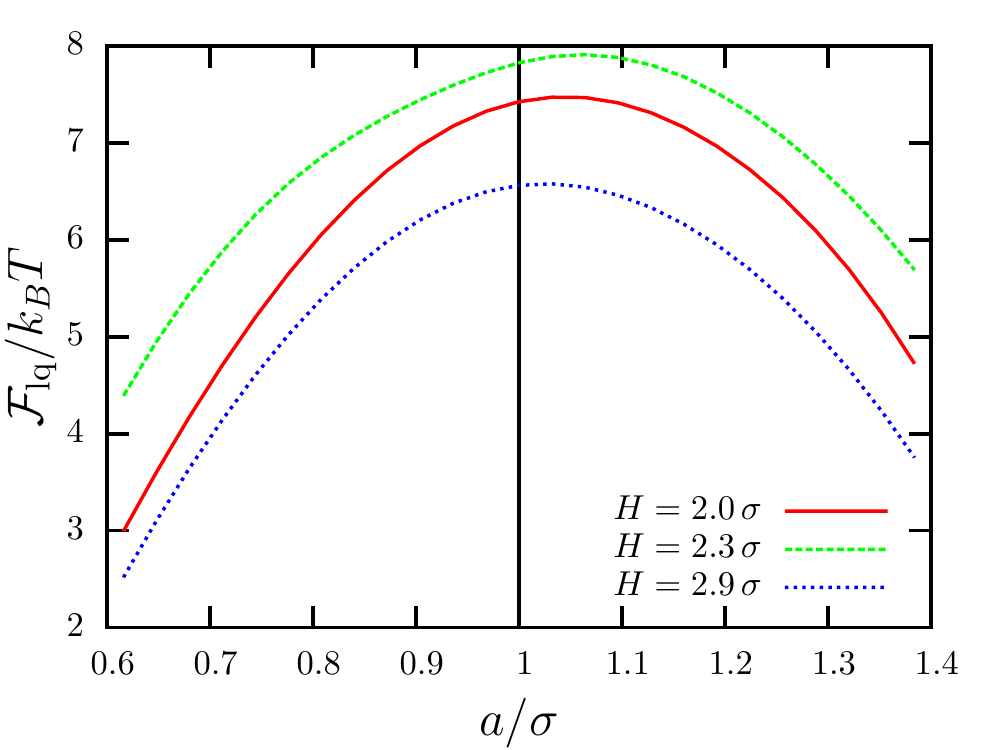}
		
		\caption{ Free energy per particle, $ \mathcal{F}_\text{lq} $ of a hard-sphere fluid in the liquid state, for particles with different diameter $ a $ in a slab with wall separation $ H $ and packing fraction $ \phi = 0.50 $. The vertical black line corresponds to the maximum of the host fluid density distribution at $ a=\sigma $.  }
		\label{fig:fmt_lq}
	\end{figure}
	The results for the free energy per particle of a hard-sphere mixture in the liquid state are shown in Fig.~\ref{fig:fmt_lq}. It can be observed that the maximum of the distribution is shifted to around $ 1.07\,\sigma $.  This is because the accessible width of the slab for hard spheres is $ L=H -a $, which means that the density of big particles (with large $ a $ and thus small $ L $) in the accessible region is higher than the one of the of small particles. This similarly holds for the ideal free energy in the liquid phase $ \mathcal{F}_\text{fl}(a) $. Furthermore there is a pronounced non-monotonic dependence of the free energy on the wall separation $ H $. In Ref.~\cite{sim:Mandal2014,sim:Mandal2017a} a similar non-monotonic behavior in the structure factor was explained with commensurate and incommensurate packing. Commensurate packing describes the favorable packing of spheres in $ k $ layers for wall separations $ H/\sigma \approx k,\,k\in \mathbb{N}  $, while incommensurate packing relates to wall separations close to half-integer values. The latter is less favorable since single particles have to be included between the $ k $ layers which also leads to a slowing-down of diffusion.
	
	\subsubsection{Cell theory}
		
		To determine the free energy of particles in the crystalline phase we employ cell theory \cite{cell:kirkwood1950,barker1963lattice,cell:Bonissent1984,theo:Schmidt1997}. As input we chose five different lattice types (2$ \triangle $, 3$ \triangle $, 3$ \astrosun $ as well as the square lattices 2$ \square $ and 3$ \square $) which can be described by two or three lattice parameters $ c_i $. One of these parameters is fixed to set the desired packing fraction $ \phi $. For each lattice type $ C $ we create a minimal structure of the crystals in a slab geometry with wall separation $ H $ and packing fraction $ \phi $ and place particles on the lattice sites with diameter $ \bar{a} $. Only one of the lattice sites is occupied with a test particle of diameter $ a $. For this particle we determine the free volume cell using a Monte Carlo procedure and thus determine its free energy based on the assumptions used in cell theory. For each $ H, \phi, \bar{a} $ and $ a $ we minimize this free energy with respect to the free lattice parameters $ c_i $. In crystals with three layers, there is a difference in the free volume cell of particles in the centre of the lattice or at the wall. In this case we determine the contributions separately and minimize the averaged free energy per particle.
		
		\subsubsection{Determination of the stability diagram}
		
		Having determined the free energies in the fluid and the crystalline phase, we assume a top-hat distribution $ f(a\,|\,\bar{a},\delta_c) $ of particles in the crystalline phase with size-dispersity $ \delta_c $ and create a list of free energy differences, 
		\begin{align}\label{key}
		\Delta \mathcal{F}&(H,\phi,C,\bar{a},\delta_c) = \Delta \mathcal{F}_\text{mix}(\delta_c) + \Delta \mathcal{F}_\text{shift} \\
		&+ \int \text{d} a \,f(a\,|\,\bar{a},\delta_c) \big[ \mathcal{F}_\text{c}(a\,|\,H,\phi,C,\bar{a})-\mathcal{F}_\text{lq}(a\,|\,H,\phi)\big].\nonumber
		\end{align} 
		Here we have to account explicitly for the mixing free energy $ \Delta \mathcal{F}_\text{mix}(\delta_c) = -\ln (\sqrt{12} \delta_c) $ of the particles in the crystal \cite{mix:Salacuse1982,mix:Sear_1998}  and include a shift parameter $ \Delta \mathcal{F}_\text{shift} $ since cell theory only provides an upper bound for the free energy. The latter was set to $ \Delta \mathcal{F}_\text{shift}=-1.80 $ in Ref.~\cite{theo:Schmidt1997} to match the value for monodisperse discs in the 2D limit. We adopted this value in our work. For every wall separation $ H $ we report the minimal packing fraction $ \phi_c $ for which the liquid and crystalline free energies are equal, $ \Delta \mathcal{F} = 0 $. Additionally, we determine the corresponding crystalline type $ C $, the mean particle diameter $ \bar{a} $ and dispersity $ \delta_c $.
		
		In a monodisperse system a coexistence region of liquid and crystalline phases can be determined from the free energies using the common tangent construction. The underlying assumption is that inside the coexistence region both the liquid and the crystalline phase have a constant packing fraction (and obviously a constant particle-size distribution). The fractionation in size-disperse systems, however, implies that the particle-size distribution in the liquid phase changes with the onset of crystallization. In Refs.~\cite{polycrist:Fasolo2003,theo:Sollich2007} is was shown that these systems can be handled using the moment free energy method, leading to highly coupled non-linear equations. In the case of confinement the application of this method is unfeasible since the free energies can only be determined numerically and it is not clear how they could be rewritten such that they only depend on four moments of the (position-dependent) density distribution. Here, we thus rely on a semi-quantitative scheme and identify the ``critical'' packing fraction $ \phi_c $ in a system with wall separation $ H $ as the packing fraction at which the free energies in the crystalline and the liquid phase are equal.

		\section{Stability diagram in slab geometries}
		\label{sec:stability}

\begin{figure}
		\includegraphics[scale=0.2]{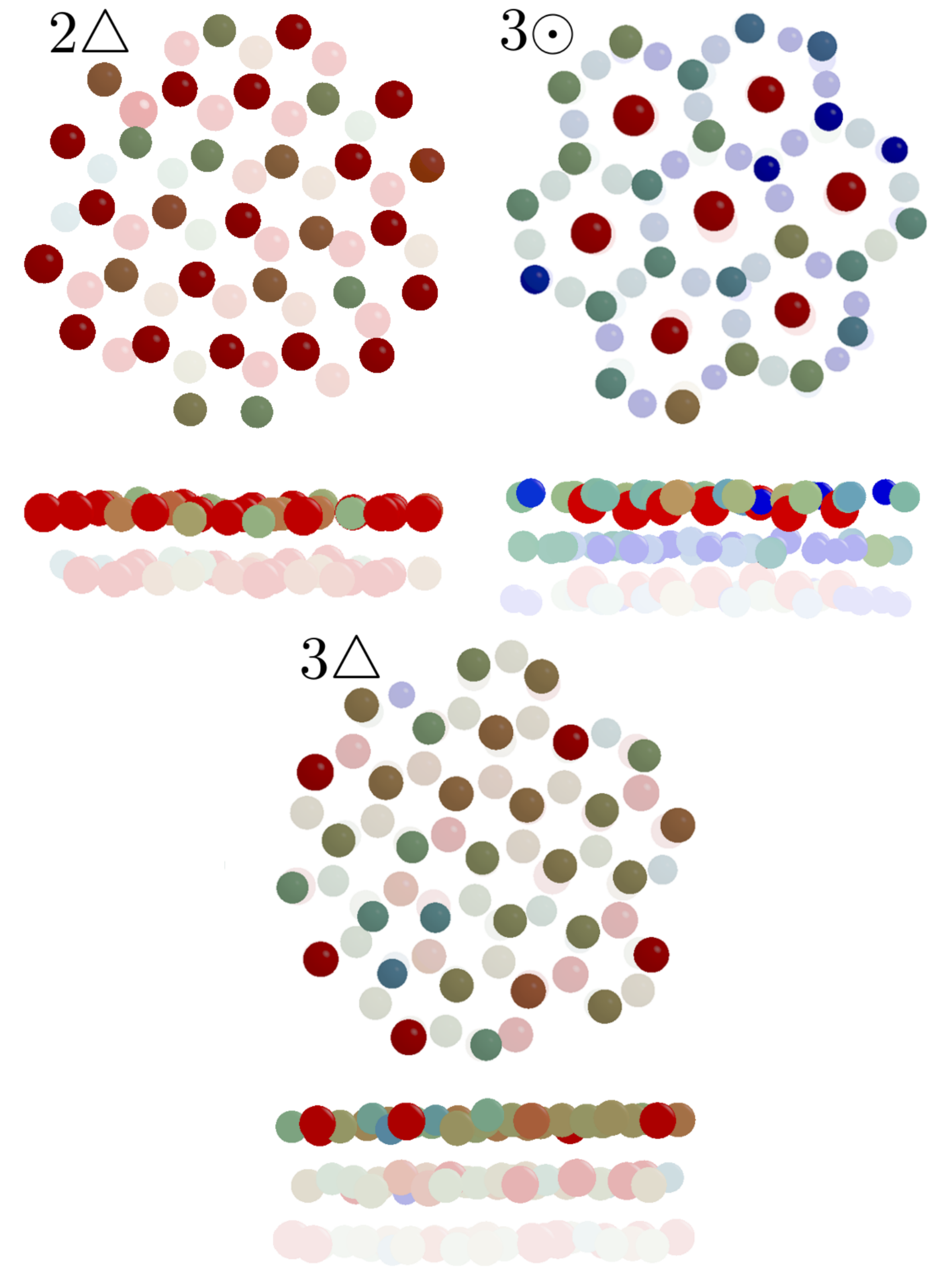}
		
		\caption{ {Visualization of the different phases observed in the slab geometry. } 2$ \triangle $) Two hexagonal layers for $ H=2.2\,\sigma, \phi=0.51 $. 3$ \astrosun $) Honeycomb phase for $ H=2.6\,\sigma, \phi=0.53 $. 3$ \triangle $) Three hexagonal layers for  $ H=3.0\,\sigma, \phi=0.53 $. Color code indicates particle diameter with $ a < 0.9\,\sigma $ (blue), $ a > 1.1\,\sigma $ (red) and interpolation between. The sphere diameter is scaled by a factor of $ 0.5 $ and the lower layers are partially transparent for the sake of visualization.  }
		\label{fig:crystal_vis}
	\end{figure}

\begin{figure}
		\hspace*{-1cm}\includegraphics[scale=0.95]{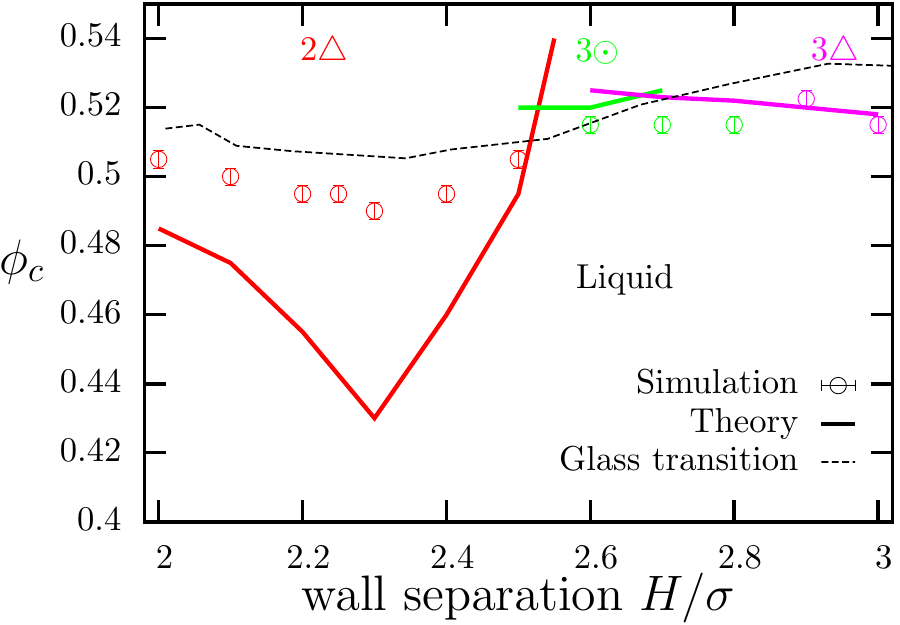}
		
		\caption{Stability diagram of size-disperse hard spheres confined between two parallel flat walls a distance $ H $ apart. Lines show theoretical and symbols simulation results for the critical packing fraction $ \phi_c $ of the liquid-to-crystal transition. Dashed line corresponds to the glass transition line of the same system as determined in Ref.~\cite{sim:Mandal2014}. Colors indicate transitions to different crystalline phases: Two hexagonal layers (2$ \triangle $, red), honeycomb phase (3$ \astrosun $, green) and three hexagonal layers (3$ \triangle $, pink). }
		\label{fig:phase_diagram}
	\end{figure}

 We first study the crystallization of hard spheres confined in slabs with different wall separations $ 2.0 < H/\sigma < 3.0. $. The system displays an onset of crystallization at packing fractions around $ \phi \approx 0.5 $, much smaller than the bulk value $ \phi_b \approx 0.6 $ \cite{polycrist:Bommineni2019}. Four different crystalline phases can be observed, namely an amorphous liquid phase for small packing fraction $ \phi $, a two-layered hexagonal structure (2$ \triangle $) for small $ H $, a three-layered hexagonal structure (3$ \triangle $) for large $ H $ and an intermediate, honeycomb-shaped phase (3$ \astrosun $). A visualization of the different phases is shown in Fig.~\ref{fig:crystal_vis}, the stability diagram is displayed in Fig.~\ref{fig:phase_diagram}. The sequence of solid-to-solid transitions $ 2\triangle\rightarrow3 \astrosun \rightarrow 3 \triangle$ upon a change of wall separation is reminiscent of similar  transitions in the case of monodisperse hard spheres or charged colloids \cite{exp:Pieranski1983,theo:Schmidt1997,exp:Winkle1986}.  Different from these monodisperse systems, the size-dispersity leads to a significantly	enlarged domain where the hexagonal structures are stable. In the small intermediate domain between $ 2\triangle $ and $ 3\triangle $ we observe the honeycomb phase, which consist of three layers of small particles forming honeycomb cells and big particles in the upper and lower layer which fill these cells, visualized in Fig.~\ref{fig:crystal_vis} (3$ \astrosun $). The formation of this honeycomb structure leads to the suppression of a square-lattice phase which was observed in the case of small or vanishing dispersity \cite{theo:Schmidt1997,Roberts2020_dyn}. In fact, for $ \delta = 0.15 $ the square-lattice is only stable in simulations at very high packing fraction $ \phi > 0.53 $ in coexistence with a liquid and a honeycomb phase. 
	
	Comparing the critical packing fractions presented in Fig.~\ref{fig:phase_diagram} to the glass transition line determined in Ref.~\cite{sim:Mandal2014} shows that the glass is metastable and can only coexist with a crystalline phase. Interestingly, both the glass transition line and the critical packing fraction $ \phi_c $ exhibit a pronounced non-monotonicity leading to the possibility of observing reentrant phase transitions by changing the wall separation \cite{reent:Bechinger1999,sim:Mandal2014}.

\begin{figure}
	 	\hspace*{-0.1cm}\includegraphics[scale=0.95]{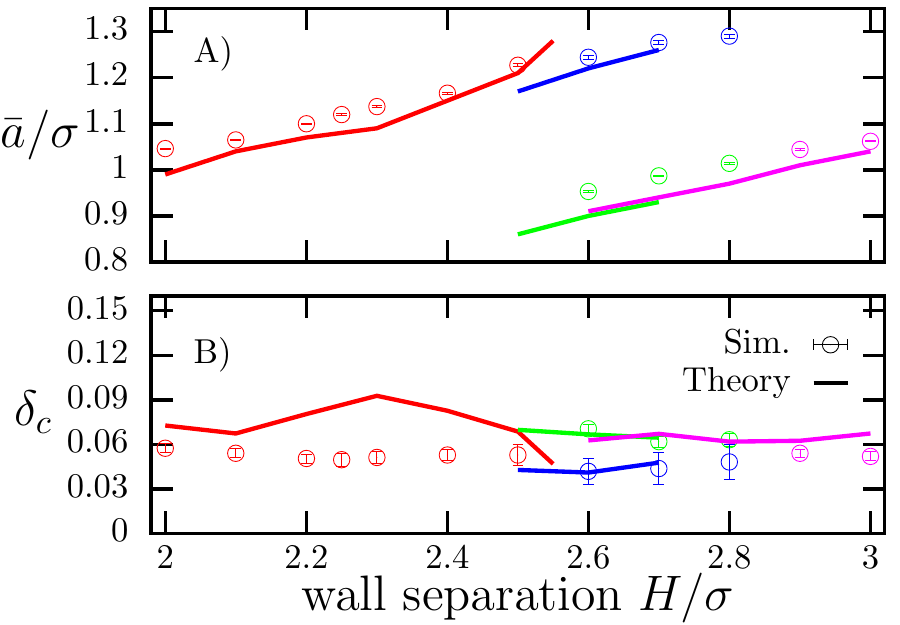}
	 	
	 	\caption{  Average particle diameter $ \bar{a} $ (A) and dispersity $ \delta_c $ (B) of particles in the different crystalline phases at packing fraction $ \phi \gtrsim \phi_c  $.  Lines show theoretical and symbols simulation results.  Colors indicate the different crystalline phases: 2$ \triangle $ (red), 3$ \astrosun $ (small particles: green, big particles: blue) and 3$ \triangle $ (pink). }
	 	\label{fig:fractionation}
	 \end{figure}

The size distribution of particles in the emergent crystals at packing fraction $ \phi \gtrsim \phi_c  $ is strongly correlated with the wall separation $ H $, as shown in Fig.~\ref{fig:fractionation}A. The average particle diameter $ \bar{a} $ for crystalline particles exhibits a linear dependence on $ H $ for one crystalline phase and a distinct jump upon a solid-to-solid transition. It can also be observed that the honeycomb phase is indeed interpolating between the two hexagonal structures since the diameters of the big and small particles in the honeycomb phase perfectly follow the linear dependence of the $ 2\triangle $ and $ 3\triangle $ hexagonal phases respectively. The size-dispersity of the crystalline particles is around $ \delta_c \approx 0.05-0.06 $, much smaller than the dispersity of the host fluid $ \delta=0.15 $, which indicates a significant fractionation (see Fig.~\ref{fig:fractionation}B). The dispersity of crystalline particles in confinement is very close to the one for crystals in bulk $ \delta_c \lesssim 0.07 $, and similar for all of the crystalline phases we observed.

	 A major impact of size-dispersity on the stability diagram  is the enlargement of domains with stable hexagonal structures compared to the monodiperse case. This is facilitated by a linear increase of the average diameter $ \bar{a} $ with wall separation $ H $, seen in Fig.~\ref{fig:fractionation}A, leading to a constant ratio $ H/\bar{a}$. Consequently, in the extreme case of a flat size-distribution of the host fluid, $ H $ would reduce to a mere length scale setting the average particle size $ \bar{a} $, leading to the existence of a single crystalline phase and a constant critical packing fraction $ \phi_c $. In reality, however, this picture is oversimplified since the particle-size distribution of the host fluid already defines the length scale $ \sigma. $ Therefore, the values of $ \bar{a} $ are bound to $ \sigma (1 - \delta) \lesssim \bar{a} \lesssim \sigma (1 + \delta)   $. Upon a further increase of $ H $ (and thus $ \bar{a} $) a discontinuous solid-to-solid transition is triggered by increasing the number of crystalline layers and reducing the average diameter of crystalline particles.

We additionally expect that the maximum in the particle-size distribution of the host fluid would lead to the existence of a minimum in $ \phi_c $ which implies a reentrent melting transition. The minimum should be approximately located at a wall separation $ H $ for which the average particle diameter $ \bar{a} = \sigma $.	Comparing Figs.~\ref{fig:phase_diagram} and \ref{fig:fractionation}B, however, shows that this minimum is shifted and placed around $ \bar{a} = 1.1\,\sigma. $ To rationalize this observation we consider the free energy distribution $ \mathcal{F}_\text{fl}(a) $ calculated with fundamental-measure theory, which indeed exhibits a shifted maximum around $ a \approx 1.07\,\sigma $ as discussed in Sec.~\ref{sec:theory}.  Additionally, the differences between commensurate and incommensurate packing in the slab leads to an increase of the liquid free energy for $ H\approx 2.5\,\sigma $ (see Fig.~\ref{fig:fmt_lq}). 
	
	Using   the semi-quantitative theory we can therefore explain the emergence and location of the reentrant melting transition. The depth of the minimum in the critical packing fraction $ \phi_c $ is, however, exaggerated in the theory, which also manifests itself in deviations from the expected dispersity in Fig.~\ref{fig:fractionation} B). That being said, the theory shows a  near-quantitative agreement with the simulation results for all investigated quantities. 
	
This analysis shows that similar to the observations in bulk \cite{polycrist:Fasolo2003,polycrist:Bommineni2019}, the crystallization of disperse particles in confinement relies on fractionation. In both cases we would expect that the free-energy barriers that have to be overcome for the particles to diffuse and order would be significantly larger compared to the case of monodisperse crystallization, which could lead to long nucleation times. For high dispersity this is likely to reach time scales much longer than typical laboratory experiments.  However, in confinement, the onset of ordering has already been observed in standard EDMD simulations (see the long wavelength peak in the structure factor for $ H=2.30\,\bar{\sigma} $  shown in Ref.~\cite{sim:Mandal2017a}).  This is likely due to the strong enhancement of heterogeneous nucleation rates compared to homogeneous nucleation \cite{hetcry:Sandomirski2011,hetcry:Espinosa2019}. Additionally, as has been discussed for bulk systems \cite{polycrist:Bommineni2019}, one possible strategy for an experimental realization is the usage of softer potentials which strongly enhance diffusion and thus speed up the crystallization \cite{exp:opal1995}. We thus believe that the described confinement-induced crystallization is also observable in laboratory experiments. The observed phenomenon is particularly important also for other applications involving external shear stresses. It has been widely observed that shear induces fractionation in channels even at moderate packing fractions, which is for example an important phenomenon for food processing \cite{shear2015}. We expect that similar fractionation effects arise in dense packings, through a combination of shear-induced crystallization \cite{shear2009} and the confinement-induced fractionation detailed here.

\section{Confinement-controlled fractionation in A wedge Geometry}
\label{sec:wedge_results}

	\begin{figure}
	\hspace*{-0.5cm}	\includegraphics[scale=0.91]{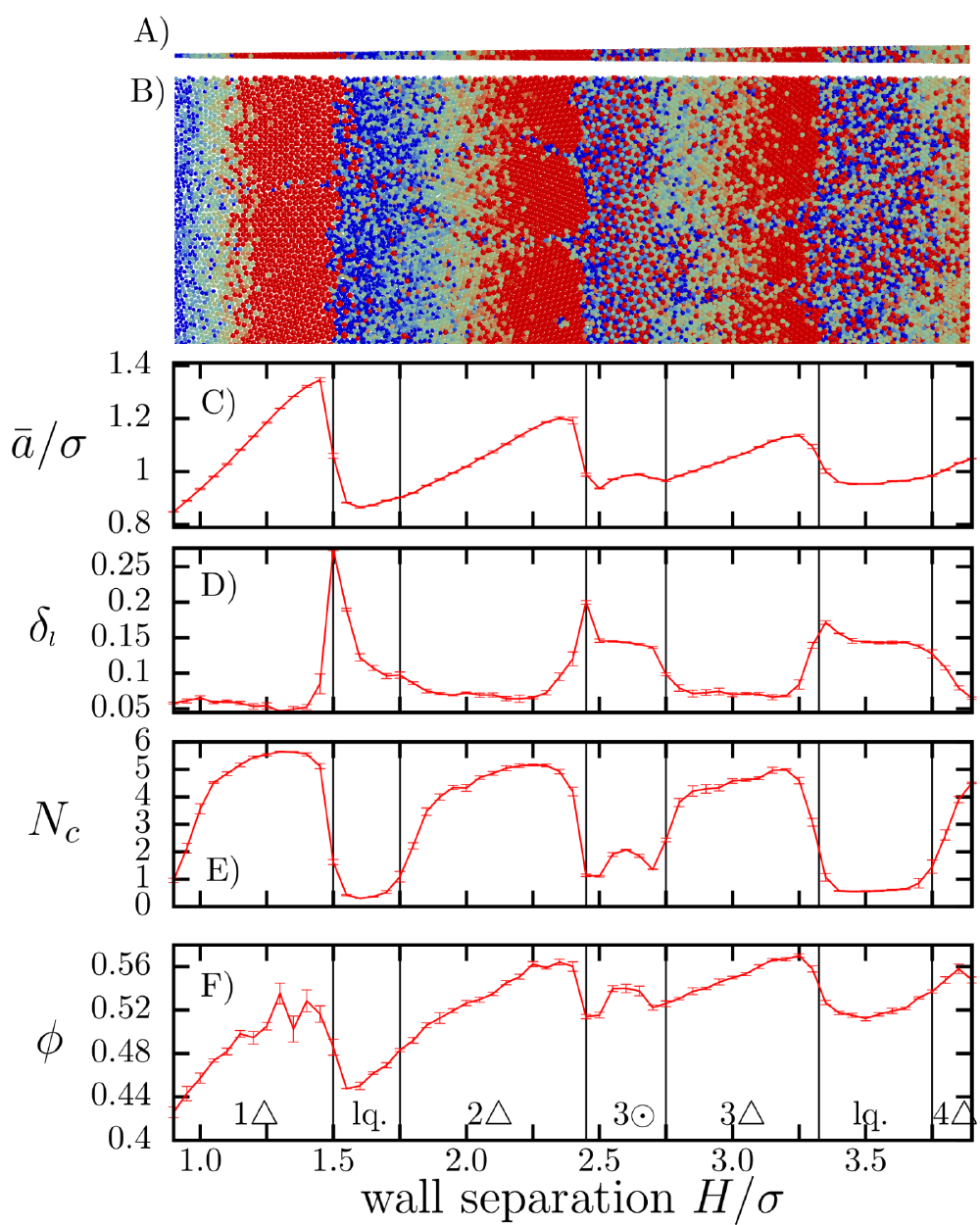}
	\caption{Demixing in a wedge geometry with average packing fraction $ \bar{\phi} = 0.51 $. Upper panels show snapshots of a typical configuration in side view (A) and top view (B). C)-F) Show local averages of mean particle diameter $ \bar{a} $, local dispersity $ \delta_l $, number of crystalline neighbors per particle $ N_\text{c} $ and packing fraction $ \phi $. The averages and errors were determined from four independent runs of the same system.  }
	\label{fig:wedge}
\end{figure}
 
We also use event-driven molecular dynamics to simulate size-disperse particles in a wedge geometry. To create the wedge one of the walls is tilted by a small angle $ \alpha $ creating a linear height profile $ H(x)=H_0 + \tan(\alpha) x $. The wedge is closed at $ H(x) = 4.0\,\sigma $ by a vertical flat wall and only the y-direction is periodic with $ L_y = 60\,\sigma. $ The wedge consists of 25,000 particles (for $ \alpha = 1^{\circ} $) leading to an average packing fraction $ \bar{\phi} = 0.51. $ We use $ N_s = 100,000 $ Monte Carlo moves per swap event and employ a similar equilibration time as used in the slab geometry. For the purpose of evaluation the wedge is separated into slices of size $ \Delta H = 0.05\,\sigma $ and local averages are calculated to determine the profiles shown in Fig.~\ref{fig:wedge}. We perform simulations for four independent wedges per tilt angle $ \alpha $ to estimate the uncertainty of the presented profiles. 

We have seen that confinement has a strong influence on the particle-size distribution. Different from the fractionation observed in bulk systems \cite{polycrist:Fasolo2003}, there is a systematic dependence of the average particle diameter on the wall separation. We can highlight this by tilting one of the walls by a small angle $ \alpha = 1^{\circ}$, thus creating a wedge geometry, seen in Fig.~\ref{fig:wedge}A \cite{sim:Mandal2014,exp:Winkle1986}. The non-parallel walls lead to a variation in the local wall separation $ H(x) $ which now depends on the x-coordinate. The average packing fraction is set to $ \bar{\phi}=0.51 $. After equilibration, separate liquid and crystal regions are observed in the wedge, displaying the same phase behaviour as discussed for the slab geometry (compare Figs.~\ref{fig:phase_diagram} and \ref{fig:wedge}B). The results are independent of the tilt angle $ \alpha $ and the inclusion of a liquid reservoir (see App.~\ref{app:wedge}). We find that the average particle diameter $ \bar{a} $ exhibits a pronounced sawtooth-like dependence on the local wall separation $ H(x) $ while the dispersity attains values between  $\delta_l\approx0.05-0.07 $ in the  crystalline regions (number of crystalline neighbors $ N_c > $ $ 3.0 $)  and $\delta_l\approx0.1-0.15 $ in the liquid regions ($ N_c < 1.0 $) (see Fig.~\ref{fig:wedge}C-E). 

We also observe large local variations in the packing fraction $ \phi(H) $ with wall separation both in the liquid and crystalline regions (see Fig.~\ref{fig:wedge}F).  Two very different factors lead to these local density variations. First, even at low average packing fraction in a pure liquid, oscillations in the local packing emerge due to the layering of the particles \cite{sim:Mandal2014}, which explains the behavior of $ \phi(H) $ in the liquid regions. Second, the local variation of the average particle diameter $ \bar{a}(H) $ in the crystalline regions leads to mechanical stress in the crystals. Assuming a uniform lattice constant we could expect a quadratic increase in the packing fraction $ \phi(H) $, since $ \bar{a} $ scales with $ H $. However, a close inspection of the snapshot in Fig.~\ref{fig:wedge}B) reveals that the system responds to the mechanical stress by bending the lattice ``planes'' (see App.~\ref{app:curvature}), which significantly reduces the variation in packing fraction. As a consequence, for small $ \bar{a} $ the packing fraction increases approximately linearly in the crystalline regions, and it is nearly constant for large $ \bar{a} $.

\section{Conclusion}
	
	We have studied the crystallization of size-disperse hard spheres in a confined geometry. The simulation results uncover solid-to-solid transitions between different crystalline phases and a non-monotonic dependence of the critical packing fraction for the liquid-to-solid transition. The latter implies the existence of a reentrant melting scenario. Moreover, we have revealed a systematic dependence of the particle-size distribution on the wall separation in the crystalline phase. To rationalize theses findings we have presented a semi-quantitative theory with which we can explain all features of the stability diagram and the emergent particle-size distributions.
	
	 Due to the universality of the underlying mechanisms which lead to the described crystallization in confinement we expect that our results generally describe the phase behaviour of particles for which excluded-volume interactions dominate. We anticipate that this holds in particular for the confinement-controlled demixing and fractionation. We have therefore devised a generally applicable technique for the purification of size-disperse particles by insertion into a wedge geometry. In the future it would be interesting to extend our studies to laboratory experiments on charged or hard-sphere-like colloids \cite{exp:Palberg2009,exp:chargedCol2016} or soft mesoscopic particles \cite{exp:opal1995}. Furthermore it would be exciting to analyze the influence of external shear stresses or different boundary conditions on the reported crystallization behavior. For example, it is expected that rough walls would significantly increase the critical packing fraction of the liquid-to-solid transition and should thus enable the existence of a stable glass phase.

	\section*{Acknowledgments}
	We acknowledge helpful discussions with Matthias Schmidt, Hartmut L\"owen, Ren\'{e} Wittmann, Suvendu Mandal, Friederike Schmid and Thomas Franosch. GJ is supported by the Austrian Science Fund (FWF): I 2887. CFP is supported by the FWF: M 2471.
	
	\appendix
	
	\section{Theoretical model: Uncertainties}
	
			\begin{figure}
	\hspace*{-1.5cm}	\includegraphics[scale=1.05]{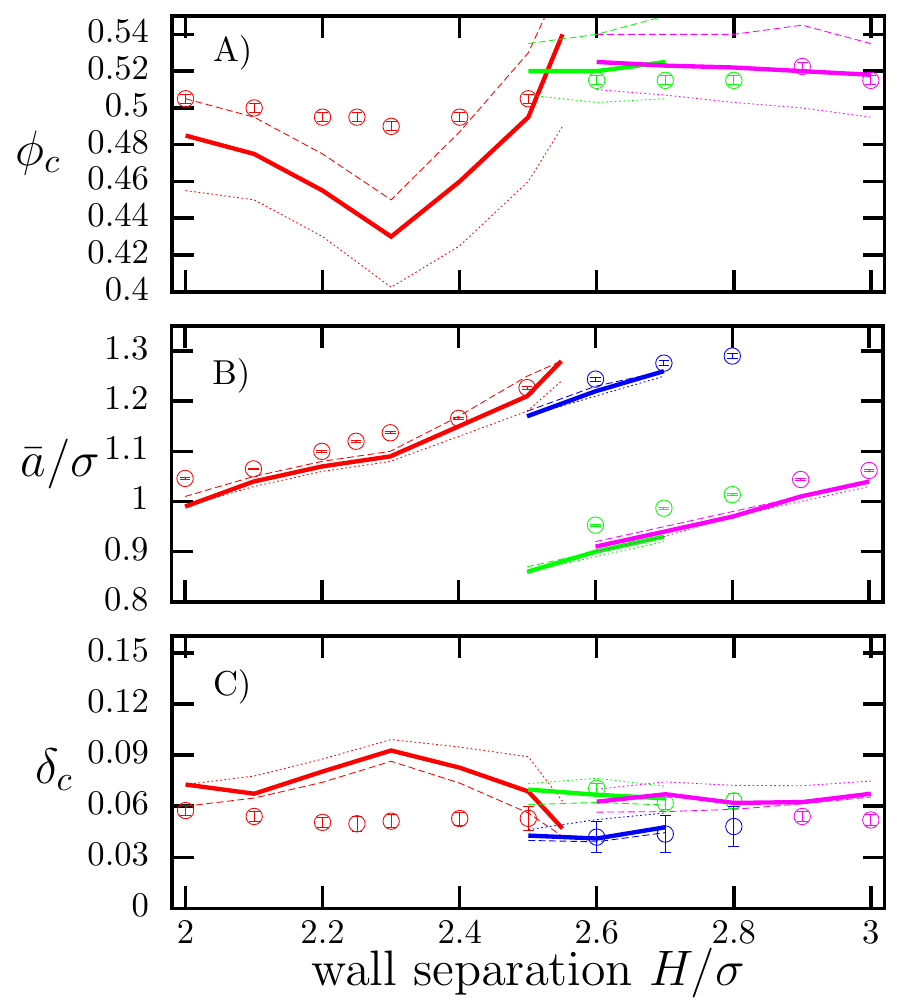}
		
		\caption{ Same as Figs.~\ref{fig:phase_diagram} and \ref{fig:fractionation} with additional lines for the error estimation of the theory. The dashed/dotted line correspond to an equivalent theory with slightly increased/decreased crystalline free energy per particle $ \Delta \mathcal{F}_\text{shift} = (- 1.80 \pm  0.25)\,k_BT $.  }
		\label{fig:combined_sm}
	\end{figure}
	
	The results for the crystallization of size-disperse hard-spheres in slab geometry are shown in Fig.~\ref{fig:combined_sm}. Additional to the data shown in Sec.~\ref{sec:stability}, the theory was also evaluated with shifted free energy differences, $ \Delta \mathcal{F}_\text{shift} = (-1.80 \pm  0.25)\,k_BT $ to estimate uncertainties originating from the assumptions behind cell theory. It should be noted that the value of $ 0.25\,k_BT $ is a lower estimate of uncertainties in the cell theory and that systematic errors can obviously not be discussed quantitatively. The most important conclusion from this analysis is that while a small shift of the free energy leads to quantitative differences, especially for the critical packing fraction, the qualitative features stay the same. It also indicates that the exaggeration of the minimum in the critical packing fraction (and the emergent maximum in the dispersity) is a systematic error of the fundamental measure theory. This error can also be revealed when comparing the density profile $ n(z) $ predicted by FMT to event-driven computer simulations as has already been discussed in Ref.~\cite{sim:Mandal2014}. Here we find that these deviations remain even without any differences in the setting up of the dispersity. We conclude that strong inhomogeneous density profiles due to broken translational symmetry are not perfectly described by FMT and lead to an overestimation of the free energy for incommensurate packing and high packing fraction. That being said, the theory shows very good agreement with simulations, despite the discussed uncontrolled approximations, and thus delivers several important explanations for the observed phenomena.
	
	\section{Dependence on tilt angle and particle reservoir in a wedge geometry}
	\label{app:wedge}
	
		\begin{figure}
		\includegraphics[scale=0.92]{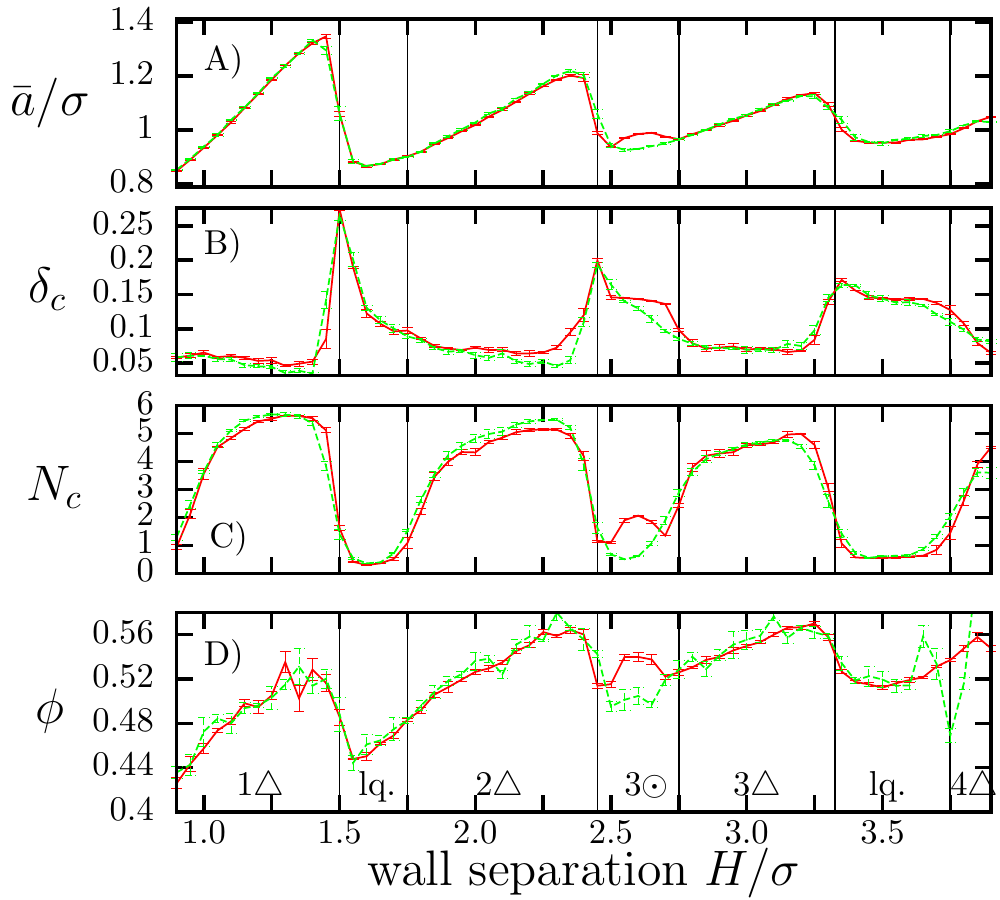}
		\caption{Demixing in a wedge geometry for average packing fraction $ \bar{\phi} = 0.51 $. The figure shows the same profiles as Fig.~\ref{fig:wedge} C)-F) for two different tilt angles $ \alpha $ ($ 1^\circ $ in red, full line and $ 2.6^\circ $ in green, dashed line).}
		\label{fig:wedge_compare}
	\end{figure}

	The resulting profiles for the wedge with $ \alpha = 1^\circ $ were presented and discussed in Sec.~\ref{sec:wedge_results}, here we compare this data to a wedge with $ \alpha = 2.6^\circ $ (see Fig.~\ref{fig:wedge_compare}). The only major difference between the profiles is the suppression of the honeycomb phase for $ \alpha=2.6^\circ $. The reason is that the region in which this phase would self-assemble becomes too small to create stable crystals. Additionally, a significant layering is observed in the packing fraction $ \phi $ at large $ H $ due to the vertical wall at $ H=4.0\,\sigma $ (see Fig.~\ref{fig:wedge_compare}D). Apart from these differences, the profiles are practically identical.
	
			\begin{figure}
		\hspace*{-1cm}\includegraphics[scale=0.91]{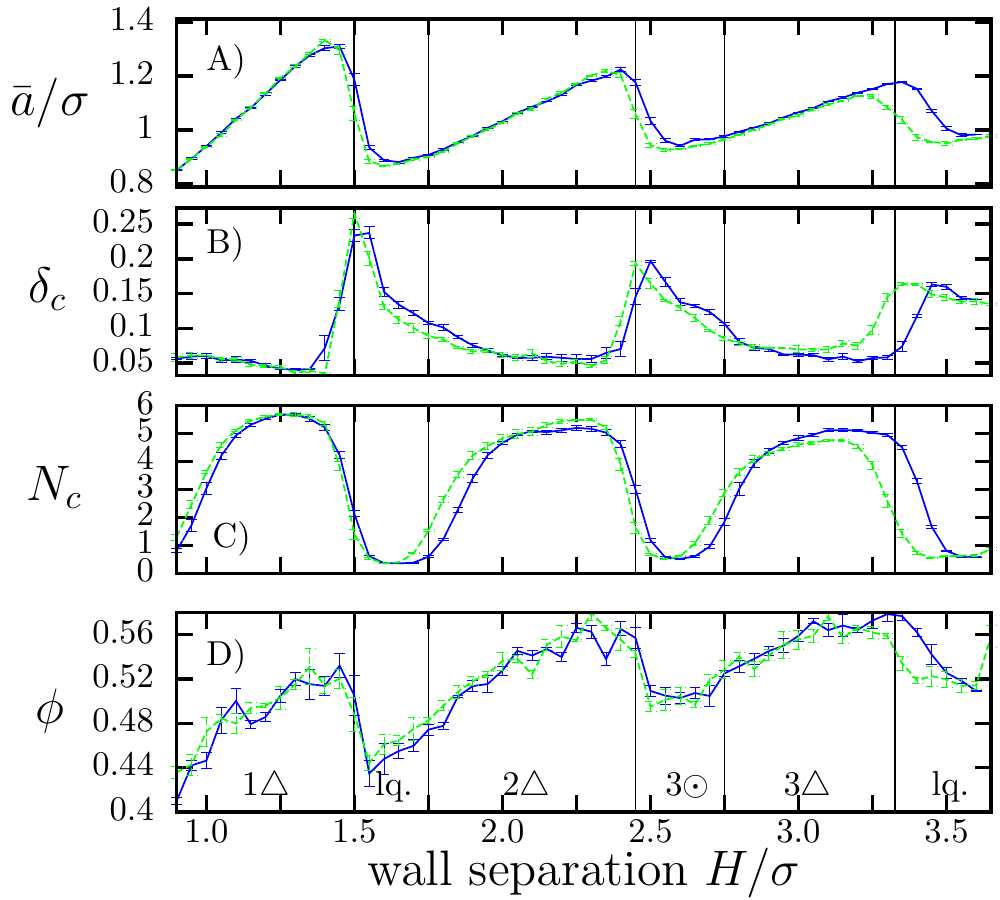}
		\caption{Demixing in a wedge geometry for average packing fraction $ \bar{\phi} = 0.51 $. The figure shows the same profiles as Fig.~\ref{fig:wedge} C)-F)  for tilt angle $ \alpha=  2.6^\circ$. The blue curves corresponds to a wedge attached to a reservoir with a slab geometry, with twice the volume of the wedge.}
		\label{fig:wedge_compare_reservoir}
	\end{figure}

	A typical laboratory experiment would correspond to a grand-canonical ensemble, because the wedge would most probably be coupled to a ``bulk'' reservoir. Due to the high packing fraction it is, however, impossible to perform a grand-canonical simulation. Here, we study the effect of the ensemble by including a reservoir in our simulations, adjacent to the wedge. This reservoir has a slab geometry with wall separation $ H=3.62\,\sigma $, and a volume twice that of the wedge.  The packing fraction was chosen such that the wedge itself has the same average packing fraction $ \bar{\phi}=0.51 $ as the ones without reservoir. The reservoir is in a liquid phase with $ \bar{a} = 0.985\,\sigma $ and $ \delta = 0.145 $ and thus sufficiently close to the particle distribution in the bulk. The results are presented in Fig.~\ref{fig:wedge_compare_reservoir}. Only marginal differences between the wedges with and without a reservoir can be observed. The reservoir increases the available number of big particles which are, as discussed before, more prone to crystallize and thus the separation effect is actually enhanced. We thus expect that the effects presented in this manuscript are independent of the ensemble and also hold in the thermodynamic limit, corresponding to a wedge with $ L_x \rightarrow \infty $.  
	
	\section{Curvature of the lattice “planes” in a wedge geometry }
		\label{app:curvature}
		
\begin{figure}[b]
	\includegraphics[keepaspectratio=true,scale=0.32]{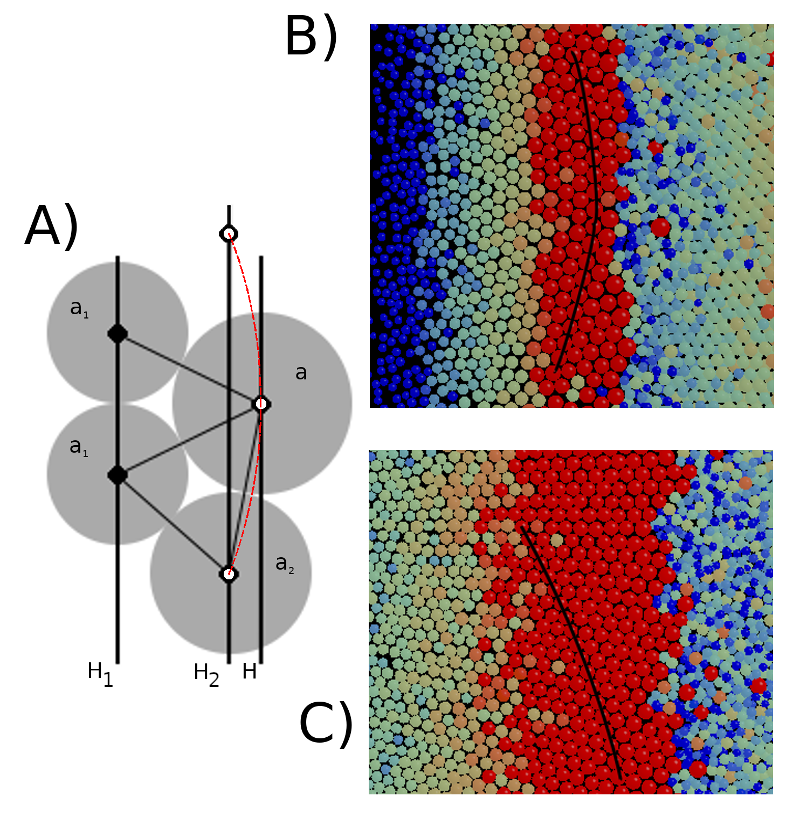}
	\caption{Curvature of the lattice ``planes'' in a wedge geometry. A) Shows a sketch of how curvature is introduced in a crystal due to local variations in the height $ H_i $ and thus the diameter $ a_i $ of the particles. The red dashed line corresponds to the circular sector with radius of curvature $ R. $ B) and C) show snapshots of EDMD simulations in wedge geometries with tilt angle $ \alpha = 0.044\,\text{rad} $ ($ 2.6^{\circ} $) and $ \alpha = 0.017\,\text{rad} $ ($ 1^{\circ} $), respectively. The black lines highlight the curved lattice planes.}\label{fig:wedge_curvature}
\end{figure}

	We also discussed in Sec.~\ref{sec:wedge_results} the role of local variations in particle diameter $ \bar{a}(H) $ inside the crystalline phases which requires the lattice ``planes'' to exhibit a certain curvature. For a one-layered hexagonal structure under the mild assumption that $ \bar{a}(x) = H(x) $ we can derive from straightforward geometrical calculations a radius of curvature, $ R $, for lattice ``planes'' in the crystalline structures. The sketch in Fig.~\ref{fig:wedge_curvature}A highlights all necessary quantities and a circular sector with the desired radius of curvature $ R $. We find,
	\begin{equation}\label{key}
	R \approx 1.1 \frac{H}{\alpha [\text{rad}]}.
	\end{equation}
	From this we can estimate the radii of curvature $ R_\alpha $ for the lattice planes marked in Fig.~\ref{fig:wedge_curvature}B,C under the assumption that $ H $ is approximately constant and find $ R_{\alpha=1^\circ} \approx 70\,\sigma $ and $ R_{\alpha=2.6^\circ} \approx 30\,\sigma $ in very good agreement with the measured values $ R^m_{\alpha=1^\circ} \approx 75.4\,\sigma $ and $ R^m_{\alpha=2.6^\circ} \approx 29.2\,\sigma $, respectively.

\bibliography{library}

\begin{thebibliography}{69}%
\makeatletter
\providecommand \@ifxundefined [1]{%
 \@ifx{#1\undefined}
}%
\providecommand \@ifnum [1]{%
 \ifnum #1\expandafter \@firstoftwo
 \else \expandafter \@secondoftwo
 \fi
}%
\providecommand \@ifx [1]{%
 \ifx #1\expandafter \@firstoftwo
 \else \expandafter \@secondoftwo
 \fi
}%
\providecommand \natexlab [1]{#1}%
\providecommand \enquote  [1]{``#1''}%
\providecommand \bibnamefont  [1]{#1}%
\providecommand \bibfnamefont [1]{#1}%
\providecommand \citenamefont [1]{#1}%
\providecommand \href@noop [0]{\@secondoftwo}%
\providecommand \href [0]{\begingroup \@sanitize@url \@href}%
\providecommand \@href[1]{\@@startlink{#1}\@@href}%
\providecommand \@@href[1]{\endgroup#1\@@endlink}%
\providecommand \@sanitize@url [0]{\catcode `\\12\catcode `\$12\catcode
  `\&12\catcode `\#12\catcode `\^12\catcode `\_12\catcode `\%12\relax}%
\providecommand \@@startlink[1]{}%
\providecommand \@@endlink[0]{}%
\providecommand \url  [0]{\begingroup\@sanitize@url \@url }%
\providecommand \@url [1]{\endgroup\@href {#1}{\urlprefix }}%
\providecommand \urlprefix  [0]{URL }%
\providecommand \Eprint [0]{\href }%
\providecommand \doibase [0]{https://doi.org/}%
\providecommand \selectlanguage [0]{\@gobble}%
\providecommand \bibinfo  [0]{\@secondoftwo}%
\providecommand \bibfield  [0]{\@secondoftwo}%
\providecommand \translation [1]{[#1]}%
\providecommand \BibitemOpen [0]{}%
\providecommand \bibitemStop [0]{}%
\providecommand \bibitemNoStop [0]{.\EOS\space}%
\providecommand \EOS [0]{\spacefactor3000\relax}%
\providecommand \BibitemShut  [1]{\csname bibitem#1\endcsname}%
\let\auto@bib@innerbib\@empty
\bibitem [{\citenamefont {Kjellander}\ and\ \citenamefont
  {Sarman}(1988)}]{theo:Kjellander1988}%
  \BibitemOpen
  \bibfield  {author} {\bibinfo {author} {\bibfnamefont {R.}~\bibnamefont
  {Kjellander}}\ and\ \bibinfo {author} {\bibfnamefont {S.}~\bibnamefont
  {Sarman}},\ }\bibfield  {title} {\bibinfo {title} {On the statistical
  mechanics of inhomogeneous fluids in narrow slits. an application to a
  hard-sphere fluid between hard walls},\ }\href
  {https://doi.org/https://doi.org/10.1016/0009-2614(88)80357-9} {\bibfield
  {journal} {\bibinfo  {journal} {Chemical Physics Letters}\ }\textbf {\bibinfo
  {volume} {149}},\ \bibinfo {pages} {102 } (\bibinfo {year}
  {1988})}\BibitemShut {NoStop}%
\bibitem [{\citenamefont {Mittal}\ \emph {et~al.}(2008)\citenamefont {Mittal},
  \citenamefont {Truskett}, \citenamefont {Errington},\ and\ \citenamefont
  {Hummer}}]{sim:Mittal2008}%
  \BibitemOpen
  \bibfield  {author} {\bibinfo {author} {\bibfnamefont {J.}~\bibnamefont
  {Mittal}}, \bibinfo {author} {\bibfnamefont {T.~M.}\ \bibnamefont
  {Truskett}}, \bibinfo {author} {\bibfnamefont {J.~R.}\ \bibnamefont
  {Errington}},\ and\ \bibinfo {author} {\bibfnamefont {G.}~\bibnamefont
  {Hummer}},\ }\bibfield  {title} {\bibinfo {title} {Layering and
  position-dependent diffusive dynamics of confined fluids},\ }\href
  {https://doi.org/10.1103/PhysRevLett.100.145901} {\bibfield  {journal}
  {\bibinfo  {journal} {Phys. Rev. Lett.}\ }\textbf {\bibinfo {volume} {100}},\
  \bibinfo {pages} {145901} (\bibinfo {year} {2008})}\BibitemShut {NoStop}%
\bibitem [{\citenamefont {G\"otzelmann}\ and\ \citenamefont
  {Dietrich}(1997)}]{theo:Gotzelmann1997}%
  \BibitemOpen
  \bibfield  {author} {\bibinfo {author} {\bibfnamefont {B.}~\bibnamefont
  {G\"otzelmann}}\ and\ \bibinfo {author} {\bibfnamefont {S.}~\bibnamefont
  {Dietrich}},\ }\bibfield  {title} {\bibinfo {title} {Density profiles and
  pair correlation functions of hard spheres in narrow slits},\ }\href
  {https://doi.org/10.1103/PhysRevE.55.2993} {\bibfield  {journal} {\bibinfo
  {journal} {Phys. Rev. E}\ }\textbf {\bibinfo {volume} {55}},\ \bibinfo
  {pages} {2993} (\bibinfo {year} {1997})}\BibitemShut {NoStop}%
\bibitem [{\citenamefont {Mandal}\ \emph {et~al.}(2017)\citenamefont {Mandal},
  \citenamefont {Lang}, \citenamefont {Boţan},\ and\ \citenamefont
  {Franosch}}]{sim:Mandal2017a}%
  \BibitemOpen
  \bibfield  {author} {\bibinfo {author} {\bibfnamefont {S.}~\bibnamefont
  {Mandal}}, \bibinfo {author} {\bibfnamefont {S.}~\bibnamefont {Lang}},
  \bibinfo {author} {\bibfnamefont {V.}~\bibnamefont {Boţan}},\ and\ \bibinfo
  {author} {\bibfnamefont {T.}~\bibnamefont {Franosch}},\ }\bibfield  {title}
  {\bibinfo {title} {{Nonergodicity parameters of confined hard-sphere
  glasses}},\ }\href {https://doi.org/10.1039/C7SM00905D} {\bibfield  {journal}
  {\bibinfo  {journal} {Soft Matter}\ }\textbf {\bibinfo {volume} {13}},\
  \bibinfo {pages} {6167} (\bibinfo {year} {2017})}\BibitemShut {NoStop}%
\bibitem [{\citenamefont {Lang}\ \emph {et~al.}(2010)\citenamefont {Lang},
  \citenamefont {Botan}, \citenamefont {Oettel}, \citenamefont {Hajnal},
  \citenamefont {Franosch},\ and\ \citenamefont {Schilling}}]{theo:Lang2010}%
  \BibitemOpen
  \bibfield  {author} {\bibinfo {author} {\bibfnamefont {S.}~\bibnamefont
  {Lang}}, \bibinfo {author} {\bibfnamefont {V.}~\bibnamefont {Botan}},
  \bibinfo {author} {\bibfnamefont {M.}~\bibnamefont {Oettel}}, \bibinfo
  {author} {\bibfnamefont {D.}~\bibnamefont {Hajnal}}, \bibinfo {author}
  {\bibfnamefont {T.}~\bibnamefont {Franosch}},\ and\ \bibinfo {author}
  {\bibfnamefont {R.}~\bibnamefont {Schilling}},\ }\bibfield  {title} {\bibinfo
  {title} {{Glass Transition in Confined Geometry}},\ }\href
  {https://doi.org/10.1103/PhysRevLett.105.125701} {\bibfield  {journal}
  {\bibinfo  {journal} {Physical Review Letters}\ }\textbf {\bibinfo {volume}
  {105}},\ \bibinfo {pages} {125701} (\bibinfo {year} {2010})}\BibitemShut
  {NoStop}%
\bibitem [{\citenamefont {Mandal}\ \emph {et~al.}(2014)\citenamefont {Mandal},
  \citenamefont {Lang}, \citenamefont {Gross}, \citenamefont {Oettel},
  \citenamefont {Raabe}, \citenamefont {Franosch},\ and\ \citenamefont
  {Varnik}}]{sim:Mandal2014}%
  \BibitemOpen
  \bibfield  {author} {\bibinfo {author} {\bibfnamefont {S.}~\bibnamefont
  {Mandal}}, \bibinfo {author} {\bibfnamefont {S.}~\bibnamefont {Lang}},
  \bibinfo {author} {\bibfnamefont {M.}~\bibnamefont {Gross}}, \bibinfo
  {author} {\bibfnamefont {M.}~\bibnamefont {Oettel}}, \bibinfo {author}
  {\bibfnamefont {D.}~\bibnamefont {Raabe}}, \bibinfo {author} {\bibfnamefont
  {T.}~\bibnamefont {Franosch}},\ and\ \bibinfo {author} {\bibfnamefont
  {F.}~\bibnamefont {Varnik}},\ }\bibfield  {title} {\bibinfo {title}
  {{Multiple reentrant glass transitions in confined hard-sphere glasses}},\
  }\href {https://doi.org/10.1038/ncomms5435} {\bibfield  {journal} {\bibinfo
  {journal} {Nature Communications}\ }\textbf {\bibinfo {volume} {5}},\
  \bibinfo {pages} {4435} (\bibinfo {year} {2014})}\BibitemShut {NoStop}%
\bibitem [{\citenamefont {Schmidt}\ and\ \citenamefont
  {L\"owen}(1996)}]{theo:Schmidt1996}%
  \BibitemOpen
  \bibfield  {author} {\bibinfo {author} {\bibfnamefont {M.}~\bibnamefont
  {Schmidt}}\ and\ \bibinfo {author} {\bibfnamefont {H.}~\bibnamefont
  {L\"owen}},\ }\bibfield  {title} {\bibinfo {title} {Freezing between two and
  three dimensions},\ }\href {https://doi.org/10.1103/PhysRevLett.76.4552}
  {\bibfield  {journal} {\bibinfo  {journal} {Phys. Rev. Lett.}\ }\textbf
  {\bibinfo {volume} {76}},\ \bibinfo {pages} {4552} (\bibinfo {year}
  {1996})}\BibitemShut {NoStop}%
\bibitem [{\citenamefont {Schmidt}\ and\ \citenamefont
  {L\"owen}(1997)}]{theo:Schmidt1997}%
  \BibitemOpen
  \bibfield  {author} {\bibinfo {author} {\bibfnamefont {M.}~\bibnamefont
  {Schmidt}}\ and\ \bibinfo {author} {\bibfnamefont {H.}~\bibnamefont
  {L\"owen}},\ }\bibfield  {title} {\bibinfo {title} {Phase diagram of hard
  spheres confined between two parallel plates},\ }\href
  {https://doi.org/10.1103/PhysRevE.55.7228} {\bibfield  {journal} {\bibinfo
  {journal} {Phys. Rev. E}\ }\textbf {\bibinfo {volume} {55}},\ \bibinfo
  {pages} {7228} (\bibinfo {year} {1997})}\BibitemShut {NoStop}%
\bibitem [{\citenamefont {Fortini}\ and\ \citenamefont
  {Dijkstra}(2006)}]{sim:Fortini_2006}%
  \BibitemOpen
  \bibfield  {author} {\bibinfo {author} {\bibfnamefont {A.}~\bibnamefont
  {Fortini}}\ and\ \bibinfo {author} {\bibfnamefont {M.}~\bibnamefont
  {Dijkstra}},\ }\bibfield  {title} {\bibinfo {title} {Phase behaviour of hard
  spheres confined between parallel hard plates: manipulation of colloidal
  crystal structures by confinement},\ }\href
  {https://doi.org/10.1088/0953-8984/18/28/l02} {\bibfield  {journal} {\bibinfo
   {journal} {Journal of Physics: Condensed Matter}\ }\textbf {\bibinfo
  {volume} {18}},\ \bibinfo {pages} {L371} (\bibinfo {year}
  {2006})}\BibitemShut {NoStop}%
\bibitem [{\citenamefont {Oguz}\ \emph {et~al.}(2012)\citenamefont {Oguz},
  \citenamefont {Marechal}, \citenamefont {Ramiro-Manzano}, \citenamefont
  {Rodriguez}, \citenamefont {Messina}, \citenamefont {Meseguer},\ and\
  \citenamefont {L\"owen}}]{sim:Oguz2012}%
  \BibitemOpen
  \bibfield  {author} {\bibinfo {author} {\bibfnamefont {E.~C.}\ \bibnamefont
  {Oguz}}, \bibinfo {author} {\bibfnamefont {M.}~\bibnamefont {Marechal}},
  \bibinfo {author} {\bibfnamefont {F.}~\bibnamefont {Ramiro-Manzano}},
  \bibinfo {author} {\bibfnamefont {I.}~\bibnamefont {Rodriguez}}, \bibinfo
  {author} {\bibfnamefont {R.}~\bibnamefont {Messina}}, \bibinfo {author}
  {\bibfnamefont {F.~J.}\ \bibnamefont {Meseguer}},\ and\ \bibinfo {author}
  {\bibfnamefont {H.}~\bibnamefont {L\"owen}},\ }\bibfield  {title} {\bibinfo
  {title} {Packing confined hard spheres denser with adaptive prism phases},\
  }\href {https://doi.org/10.1103/PhysRevLett.109.218301} {\bibfield  {journal}
  {\bibinfo  {journal} {Phys. Rev. Lett.}\ }\textbf {\bibinfo {volume} {109}},\
  \bibinfo {pages} {218301} (\bibinfo {year} {2012})}\BibitemShut {NoStop}%
\bibitem [{\citenamefont {Scheidler}\ \emph {et~al.}(2002)\citenamefont
  {Scheidler}, \citenamefont {Kob},\ and\ \citenamefont
  {Binder}}]{sim:Scheidler2002}%
  \BibitemOpen
  \bibfield  {author} {\bibinfo {author} {\bibfnamefont {P.}~\bibnamefont
  {Scheidler}}, \bibinfo {author} {\bibfnamefont {W.}~\bibnamefont {Kob}},\
  and\ \bibinfo {author} {\bibfnamefont {K.}~\bibnamefont {Binder}},\
  }\bibfield  {title} {\bibinfo {title} {Cooperative motion and growing length
  scales in supercooled confined liquids},\ }\href
  {https://doi.org/10.1209/epl/i2002-00182-9} {\bibfield  {journal} {\bibinfo
  {journal} {Europhysics Letters ({EPL})}\ }\textbf {\bibinfo {volume} {59}},\
  \bibinfo {pages} {701} (\bibinfo {year} {2002})}\BibitemShut {NoStop}%
\bibitem [{\citenamefont {Roberts}\ \emph {et~al.}(2020)\citenamefont
  {Roberts}, \citenamefont {Marioni}, \citenamefont {Palmer},\ and\
  \citenamefont {Conrad}}]{Roberts2020_dyn}%
  \BibitemOpen
  \bibfield  {author} {\bibinfo {author} {\bibfnamefont {R.~C.}\ \bibnamefont
  {Roberts}}, \bibinfo {author} {\bibfnamefont {N.}~\bibnamefont {Marioni}},
  \bibinfo {author} {\bibfnamefont {J.~C.}\ \bibnamefont {Palmer}},\ and\
  \bibinfo {author} {\bibfnamefont {J.~C.}\ \bibnamefont {Conrad}},\ }\bibfield
   {title} {\bibinfo {title} {Dynamics of polydisperse hard-spheres under
  strong confinement},\ }\href {https://doi.org/10.1080/00268976.2020.1728407}
  {\bibfield  {journal} {\bibinfo  {journal} {Molecular Physics}\ }\textbf
  {\bibinfo {volume} {0}},\ \bibinfo {pages} {1} (\bibinfo {year}
  {2020})}\BibitemShut {NoStop}%
\bibitem [{\citenamefont {Nugent}\ \emph {et~al.}(2007)\citenamefont {Nugent},
  \citenamefont {Edmond}, \citenamefont {Patel},\ and\ \citenamefont
  {Weeks}}]{exp:Nugent2007}%
  \BibitemOpen
  \bibfield  {author} {\bibinfo {author} {\bibfnamefont {C.~R.}\ \bibnamefont
  {Nugent}}, \bibinfo {author} {\bibfnamefont {K.~V.}\ \bibnamefont {Edmond}},
  \bibinfo {author} {\bibfnamefont {H.~N.}\ \bibnamefont {Patel}},\ and\
  \bibinfo {author} {\bibfnamefont {E.~R.}\ \bibnamefont {Weeks}},\ }\bibfield
  {title} {\bibinfo {title} {Colloidal glass transition observed in
  confinement},\ }\href {https://doi.org/10.1103/PhysRevLett.99.025702}
  {\bibfield  {journal} {\bibinfo  {journal} {Phys. Rev. Lett.}\ }\textbf
  {\bibinfo {volume} {99}},\ \bibinfo {pages} {025702} (\bibinfo {year}
  {2007})}\BibitemShut {NoStop}%
\bibitem [{\citenamefont {Sarangapani}\ \emph {et~al.}(2011)\citenamefont
  {Sarangapani}, \citenamefont {Schofield},\ and\ \citenamefont
  {Zhu}}]{exp:Sarangapani2011}%
  \BibitemOpen
  \bibfield  {author} {\bibinfo {author} {\bibfnamefont {P.~S.}\ \bibnamefont
  {Sarangapani}}, \bibinfo {author} {\bibfnamefont {A.~B.}\ \bibnamefont
  {Schofield}},\ and\ \bibinfo {author} {\bibfnamefont {Y.}~\bibnamefont
  {Zhu}},\ }\bibfield  {title} {\bibinfo {title} {Direct experimental evidence
  of growing dynamic length scales in confined colloidal liquids},\ }\href
  {https://doi.org/10.1103/PhysRevE.83.030502} {\bibfield  {journal} {\bibinfo
  {journal} {Phys. Rev. E}\ }\textbf {\bibinfo {volume} {83}},\ \bibinfo
  {pages} {030502(R)} (\bibinfo {year} {2011})}\BibitemShut {NoStop}%
\bibitem [{\citenamefont {Reinmüller}\ \emph {et~al.}(2012)\citenamefont
  {Reinmüller}, \citenamefont {Oğuz}, \citenamefont {Messina}, \citenamefont
  {Löwen}, \citenamefont {Schöpe},\ and\ \citenamefont
  {Palberg}}]{exp:Reinmuller2012}%
  \BibitemOpen
  \bibfield  {author} {\bibinfo {author} {\bibfnamefont {A.}~\bibnamefont
  {Reinmüller}}, \bibinfo {author} {\bibfnamefont {E.~C.}\ \bibnamefont
  {Oğuz}}, \bibinfo {author} {\bibfnamefont {R.}~\bibnamefont {Messina}},
  \bibinfo {author} {\bibfnamefont {H.}~\bibnamefont {Löwen}}, \bibinfo
  {author} {\bibfnamefont {H.~J.}\ \bibnamefont {Schöpe}},\ and\ \bibinfo
  {author} {\bibfnamefont {T.}~\bibnamefont {Palberg}},\ }\bibfield  {title}
  {\bibinfo {title} {Colloidal crystallization in the quasi-two-dimensional
  induced by electrolyte gradients},\ }\href
  {https://doi.org/10.1063/1.4705393} {\bibfield  {journal} {\bibinfo
  {journal} {The Journal of Chemical Physics}\ }\textbf {\bibinfo {volume}
  {136}},\ \bibinfo {pages} {164505} (\bibinfo {year} {2012})}\BibitemShut
  {NoStop}%
\bibitem [{\citenamefont {Nyg\aa{}rd}\ \emph {et~al.}(2012)\citenamefont
  {Nyg\aa{}rd}, \citenamefont {Kjellander}, \citenamefont {Sarman},
  \citenamefont {Chodankar}, \citenamefont {Perret}, \citenamefont
  {Buitenhuis},\ and\ \citenamefont {van~der Veen}}]{exp:nygard2012}%
  \BibitemOpen
  \bibfield  {author} {\bibinfo {author} {\bibfnamefont {K.}~\bibnamefont
  {Nyg\aa{}rd}}, \bibinfo {author} {\bibfnamefont {R.}~\bibnamefont
  {Kjellander}}, \bibinfo {author} {\bibfnamefont {S.}~\bibnamefont {Sarman}},
  \bibinfo {author} {\bibfnamefont {S.}~\bibnamefont {Chodankar}}, \bibinfo
  {author} {\bibfnamefont {E.}~\bibnamefont {Perret}}, \bibinfo {author}
  {\bibfnamefont {J.}~\bibnamefont {Buitenhuis}},\ and\ \bibinfo {author}
  {\bibfnamefont {J.~F.}\ \bibnamefont {van~der Veen}},\ }\bibfield  {title}
  {\bibinfo {title} {Anisotropic pair correlations and structure factors of
  confined hard-sphere fluids: An experimental and theoretical study},\ }\href
  {https://doi.org/10.1103/PhysRevLett.108.037802} {\bibfield  {journal}
  {\bibinfo  {journal} {Phys. Rev. Lett.}\ }\textbf {\bibinfo {volume} {108}},\
  \bibinfo {pages} {037802} (\bibinfo {year} {2012})}\BibitemShut {NoStop}%
\bibitem [{\citenamefont {Nygård}\ \emph {et~al.}(2013)\citenamefont
  {Nygård}, \citenamefont {Sarman},\ and\ \citenamefont
  {Kjellander}}]{exp:nygard2013}%
  \BibitemOpen
  \bibfield  {author} {\bibinfo {author} {\bibfnamefont {K.}~\bibnamefont
  {Nygård}}, \bibinfo {author} {\bibfnamefont {S.}~\bibnamefont {Sarman}},\
  and\ \bibinfo {author} {\bibfnamefont {R.}~\bibnamefont {Kjellander}},\
  }\bibfield  {title} {\bibinfo {title} {Local order variations in confined
  hard-sphere fluids},\ }\href {https://doi.org/10.1063/1.4825176} {\bibfield
  {journal} {\bibinfo  {journal} {The Journal of Chemical Physics}\ }\textbf
  {\bibinfo {volume} {139}},\ \bibinfo {pages} {164701} (\bibinfo {year}
  {2013})}\BibitemShut {NoStop}%
\bibitem [{\citenamefont {Hunter}\ \emph {et~al.}(2014)\citenamefont {Hunter},
  \citenamefont {Edmond},\ and\ \citenamefont {Weeks}}]{exp:Hunter2014}%
  \BibitemOpen
  \bibfield  {author} {\bibinfo {author} {\bibfnamefont {G.~L.}\ \bibnamefont
  {Hunter}}, \bibinfo {author} {\bibfnamefont {K.~V.}\ \bibnamefont {Edmond}},\
  and\ \bibinfo {author} {\bibfnamefont {E.~R.}\ \bibnamefont {Weeks}},\
  }\bibfield  {title} {\bibinfo {title} {Boundary mobility controls glassiness
  in confined colloidal liquids},\ }\href
  {https://doi.org/10.1103/PhysRevLett.112.218302} {\bibfield  {journal}
  {\bibinfo  {journal} {Phys. Rev. Lett.}\ }\textbf {\bibinfo {volume} {112}},\
  \bibinfo {pages} {218302} (\bibinfo {year} {2014})}\BibitemShut {NoStop}%
\bibitem [{\citenamefont {Williams}\ \emph {et~al.}(2015)\citenamefont
  {Williams}, \citenamefont {Oğuz}, \citenamefont {Bartlett}, \citenamefont
  {Löwen},\ and\ \citenamefont {Patrick~Royall}}]{exp:Williams2015}%
  \BibitemOpen
  \bibfield  {author} {\bibinfo {author} {\bibfnamefont {I.}~\bibnamefont
  {Williams}}, \bibinfo {author} {\bibfnamefont {E.~C.}\ \bibnamefont {Oğuz}},
  \bibinfo {author} {\bibfnamefont {P.}~\bibnamefont {Bartlett}}, \bibinfo
  {author} {\bibfnamefont {H.}~\bibnamefont {Löwen}},\ and\ \bibinfo {author}
  {\bibfnamefont {C.}~\bibnamefont {Patrick~Royall}},\ }\bibfield  {title}
  {\bibinfo {title} {Flexible confinement leads to multiple relaxation regimes
  in glassy colloidal liquids},\ }\href {https://doi.org/10.1063/1.4905472}
  {\bibfield  {journal} {\bibinfo  {journal} {The Journal of Chemical Physics}\
  }\textbf {\bibinfo {volume} {142}},\ \bibinfo {pages} {024505} (\bibinfo
  {year} {2015})}\BibitemShut {NoStop}%
\bibitem [{\citenamefont {Nyg\aa{}rd}\ \emph {et~al.}(2016)\citenamefont
  {Nyg\aa{}rd}, \citenamefont {Sarman}, \citenamefont {Hyltegren},
  \citenamefont {Chodankar}, \citenamefont {Perret}, \citenamefont
  {Buitenhuis}, \citenamefont {van~der Veen},\ and\ \citenamefont
  {Kjellander}}]{exp:nygard2016_2}%
  \BibitemOpen
  \bibfield  {author} {\bibinfo {author} {\bibfnamefont {K.}~\bibnamefont
  {Nyg\aa{}rd}}, \bibinfo {author} {\bibfnamefont {S.}~\bibnamefont {Sarman}},
  \bibinfo {author} {\bibfnamefont {K.}~\bibnamefont {Hyltegren}}, \bibinfo
  {author} {\bibfnamefont {S.}~\bibnamefont {Chodankar}}, \bibinfo {author}
  {\bibfnamefont {E.}~\bibnamefont {Perret}}, \bibinfo {author} {\bibfnamefont
  {J.}~\bibnamefont {Buitenhuis}}, \bibinfo {author} {\bibfnamefont {J.~F.}\
  \bibnamefont {van~der Veen}},\ and\ \bibinfo {author} {\bibfnamefont
  {R.}~\bibnamefont {Kjellander}},\ }\bibfield  {title} {\bibinfo {title}
  {Density fluctuations of hard-sphere fluids in narrow confinement},\ }\href
  {https://doi.org/10.1103/PhysRevX.6.011014} {\bibfield  {journal} {\bibinfo
  {journal} {Phys. Rev. X}\ }\textbf {\bibinfo {volume} {6}},\ \bibinfo {pages}
  {011014} (\bibinfo {year} {2016})}\BibitemShut {NoStop}%
\bibitem [{\citenamefont {Zhang}\ and\ \citenamefont
  {Cheng}(2016)}]{exp:Zhang2016}%
  \BibitemOpen
  \bibfield  {author} {\bibinfo {author} {\bibfnamefont {B.}~\bibnamefont
  {Zhang}}\ and\ \bibinfo {author} {\bibfnamefont {X.}~\bibnamefont {Cheng}},\
  }\bibfield  {title} {\bibinfo {title} {Structures and dynamics of
  glass-forming colloidal liquids under spherical confinement},\ }\href
  {https://doi.org/10.1103/PhysRevLett.116.098302} {\bibfield  {journal}
  {\bibinfo  {journal} {Phys. Rev. Lett.}\ }\textbf {\bibinfo {volume} {116}},\
  \bibinfo {pages} {098302} (\bibinfo {year} {2016})}\BibitemShut {NoStop}%
\bibitem [{\citenamefont {Kienle}\ and\ \citenamefont
  {Kuhl}(2016)}]{exp:Kienle2016}%
  \BibitemOpen
  \bibfield  {author} {\bibinfo {author} {\bibfnamefont {D.~F.}\ \bibnamefont
  {Kienle}}\ and\ \bibinfo {author} {\bibfnamefont {T.~L.}\ \bibnamefont
  {Kuhl}},\ }\bibfield  {title} {\bibinfo {title} {Density and phase state of a
  confined nonpolar fluid},\ }\href
  {https://doi.org/10.1103/PhysRevLett.117.036101} {\bibfield  {journal}
  {\bibinfo  {journal} {Phys. Rev. Lett.}\ }\textbf {\bibinfo {volume} {117}},\
  \bibinfo {pages} {036101} (\bibinfo {year} {2016})}\BibitemShut {NoStop}%
\bibitem [{\citenamefont {Lippmann}\ \emph {et~al.}(2019)\citenamefont
  {Lippmann}, \citenamefont {Seeck}, \citenamefont {Ehnes}, \citenamefont
  {Nygård}, \citenamefont {Bertram},\ and\ \citenamefont
  {Ciobanu}}]{exp:Lippmann2019}%
  \BibitemOpen
  \bibfield  {author} {\bibinfo {author} {\bibfnamefont {M.}~\bibnamefont
  {Lippmann}}, \bibinfo {author} {\bibfnamefont {O.~H.}\ \bibnamefont {Seeck}},
  \bibinfo {author} {\bibfnamefont {A.}~\bibnamefont {Ehnes}}, \bibinfo
  {author} {\bibfnamefont {K.}~\bibnamefont {Nygård}}, \bibinfo {author}
  {\bibfnamefont {F.}~\bibnamefont {Bertram}},\ and\ \bibinfo {author}
  {\bibfnamefont {A.}~\bibnamefont {Ciobanu}},\ }\bibfield  {title} {\bibinfo
  {title} {Experimental observation of crystal–liquid coexistence in
  slit-confined nonpolar fluids},\ }\href
  {https://doi.org/10.1021/acs.jpclett.9b00331} {\bibfield  {journal} {\bibinfo
   {journal} {The Journal of Physical Chemistry Letters}\ }\textbf {\bibinfo
  {volume} {10}},\ \bibinfo {pages} {1634} (\bibinfo {year}
  {2019})}\BibitemShut {NoStop}%
\bibitem [{\citenamefont {Rosenfeld}\ \emph {et~al.}(1997)\citenamefont
  {Rosenfeld}, \citenamefont {Schmidt}, \citenamefont {L\"owen},\ and\
  \citenamefont {Tarazona}}]{theo:Rosenfeld1997}%
  \BibitemOpen
  \bibfield  {author} {\bibinfo {author} {\bibfnamefont {Y.}~\bibnamefont
  {Rosenfeld}}, \bibinfo {author} {\bibfnamefont {M.}~\bibnamefont {Schmidt}},
  \bibinfo {author} {\bibfnamefont {H.}~\bibnamefont {L\"owen}},\ and\ \bibinfo
  {author} {\bibfnamefont {P.}~\bibnamefont {Tarazona}},\ }\bibfield  {title}
  {\bibinfo {title} {Fundamental-measure free-energy density functional for
  hard spheres: Dimensional crossover and freezing},\ }\href
  {https://doi.org/10.1103/PhysRevE.55.4245} {\bibfield  {journal} {\bibinfo
  {journal} {Phys. Rev. E}\ }\textbf {\bibinfo {volume} {55}},\ \bibinfo
  {pages} {4245} (\bibinfo {year} {1997})}\BibitemShut {NoStop}%
\bibitem [{\citenamefont {Grodon}\ \emph {et~al.}(2005)\citenamefont {Grodon},
  \citenamefont {Dijkstra}, \citenamefont {Evans},\ and\ \citenamefont
  {Roth}}]{theo:Roth2005}%
  \BibitemOpen
  \bibfield  {author} {\bibinfo {author} {\bibfnamefont {C.}~\bibnamefont
  {Grodon}}, \bibinfo {author} {\bibfnamefont {M.}~\bibnamefont {Dijkstra}},
  \bibinfo {author} {\bibfnamefont {R.}~\bibnamefont {Evans}},\ and\ \bibinfo
  {author} {\bibfnamefont {R.}~\bibnamefont {Roth}},\ }\bibfield  {title}
  {\bibinfo {title} {Homogeneous and inhomogeneous hard-sphere mixtures:
  manifestations of structural crossover},\ }\href
  {https://doi.org/10.1080/00268970500167532} {\bibfield  {journal} {\bibinfo
  {journal} {Molecular Physics}\ }\textbf {\bibinfo {volume} {103}},\ \bibinfo
  {pages} {3009} (\bibinfo {year} {2005})},\ \Eprint
  {https://arxiv.org/abs/https://doi.org/10.1080/00268970500167532}
  {https://doi.org/10.1080/00268970500167532} \BibitemShut {NoStop}%
\bibitem [{\citenamefont {O{\u{g}}uz}\ \emph {et~al.}(2009)\citenamefont
  {O{\u{g}}uz}, \citenamefont {Messina},\ and\ \citenamefont
  {Löwen}}]{theo:Oguz2009}%
  \BibitemOpen
  \bibfield  {author} {\bibinfo {author} {\bibfnamefont {E.~C.}\ \bibnamefont
  {O{\u{g}}uz}}, \bibinfo {author} {\bibfnamefont {R.}~\bibnamefont
  {Messina}},\ and\ \bibinfo {author} {\bibfnamefont {H.}~\bibnamefont
  {Löwen}},\ }\bibfield  {title} {\bibinfo {title} {Multilayered crystals of
  macroions under slit confinement},\ }\href
  {https://doi.org/10.1088/0953-8984/21/42/424110} {\bibfield  {journal}
  {\bibinfo  {journal} {Journal of Physics: Condensed Matter}\ }\textbf
  {\bibinfo {volume} {21}},\ \bibinfo {pages} {424110} (\bibinfo {year}
  {2009})}\BibitemShut {NoStop}%
\bibitem [{\citenamefont {Fehr}\ and\ \citenamefont
  {L\"owen}(1995)}]{sim:Fehr1995}%
  \BibitemOpen
  \bibfield  {author} {\bibinfo {author} {\bibfnamefont {T.}~\bibnamefont
  {Fehr}}\ and\ \bibinfo {author} {\bibfnamefont {H.}~\bibnamefont {L\"owen}},\
  }\bibfield  {title} {\bibinfo {title} {Glass transition in confined
  geometry},\ }\href {https://doi.org/10.1103/PhysRevE.52.4016} {\bibfield
  {journal} {\bibinfo  {journal} {Phys. Rev. E}\ }\textbf {\bibinfo {volume}
  {52}},\ \bibinfo {pages} {4016} (\bibinfo {year} {1995})}\BibitemShut
  {NoStop}%
\bibitem [{\citenamefont {Rodríguez}\ and\ \citenamefont
  {Vicente}(1996)}]{sim:Rodriguez1996}%
  \BibitemOpen
  \bibfield  {author} {\bibinfo {author} {\bibfnamefont {G.}~\bibnamefont
  {Rodríguez}}\ and\ \bibinfo {author} {\bibfnamefont {L.}~\bibnamefont
  {Vicente}},\ }\bibfield  {title} {\bibinfo {title} {Density profiles of
  colloidal suspensions in equilibrium inside slit pores},\ }\href
  {https://doi.org/10.1080/00268979600100121} {\bibfield  {journal} {\bibinfo
  {journal} {Molecular Physics}\ }\textbf {\bibinfo {volume} {87}},\ \bibinfo
  {pages} {213} (\bibinfo {year} {1996})}\BibitemShut {NoStop}%
\bibitem [{\citenamefont {Mittal}\ \emph {et~al.}(2007)\citenamefont {Mittal},
  \citenamefont {Errington},\ and\ \citenamefont
  {Truskett}}]{sim:Mittal2007_2}%
  \BibitemOpen
  \bibfield  {author} {\bibinfo {author} {\bibfnamefont {J.}~\bibnamefont
  {Mittal}}, \bibinfo {author} {\bibfnamefont {J.~R.}\ \bibnamefont
  {Errington}},\ and\ \bibinfo {author} {\bibfnamefont {T.~M.}\ \bibnamefont
  {Truskett}},\ }\bibfield  {title} {\bibinfo {title} {Relationships between
  self-diffusivity, packing fraction, and excess entropy in simple bulk and
  confined fluids},\ }\href {https://doi.org/10.1021/jp071369e} {\bibfield
  {journal} {\bibinfo  {journal} {The Journal of Physical Chemistry B}\
  }\textbf {\bibinfo {volume} {111}},\ \bibinfo {pages} {10054} (\bibinfo
  {year} {2007})},\ \bibinfo {note} {pMID: 17629320}\BibitemShut {NoStop}%
\bibitem [{\citenamefont {Goel}\ \emph {et~al.}(2009)\citenamefont {Goel},
  \citenamefont {Krekelberg}, \citenamefont {Pond}, \citenamefont {Mittal},
  \citenamefont {Shen}, \citenamefont {Errington},\ and\ \citenamefont
  {Truskett}}]{sim:Goel2009}%
  \BibitemOpen
  \bibfield  {author} {\bibinfo {author} {\bibfnamefont {G.}~\bibnamefont
  {Goel}}, \bibinfo {author} {\bibfnamefont {W.~P.}\ \bibnamefont
  {Krekelberg}}, \bibinfo {author} {\bibfnamefont {M.~J.}\ \bibnamefont
  {Pond}}, \bibinfo {author} {\bibfnamefont {J.}~\bibnamefont {Mittal}},
  \bibinfo {author} {\bibfnamefont {V.~K.}\ \bibnamefont {Shen}}, \bibinfo
  {author} {\bibfnamefont {J.~R.}\ \bibnamefont {Errington}},\ and\ \bibinfo
  {author} {\bibfnamefont {T.~M.}\ \bibnamefont {Truskett}},\ }\bibfield
  {title} {\bibinfo {title} {Available states and available space: static
  properties that predict self-diffusivity of confined fluids},\ }\href
  {https://doi.org/10.1088/1742-5468/2009/04/p04006} {\bibfield  {journal}
  {\bibinfo  {journal} {Journal of Statistical Mechanics: Theory and
  Experiment}\ }\textbf {\bibinfo {volume} {2009}},\ \bibinfo {pages} {P04006}
  (\bibinfo {year} {2009})}\BibitemShut {NoStop}%
\bibitem [{\citenamefont {Krishnan}\ and\ \citenamefont
  {Ayappa}(2012)}]{sim:Krishnan2012}%
  \BibitemOpen
  \bibfield  {author} {\bibinfo {author} {\bibfnamefont {S.~H.}\ \bibnamefont
  {Krishnan}}\ and\ \bibinfo {author} {\bibfnamefont {K.~G.}\ \bibnamefont
  {Ayappa}},\ }\bibfield  {title} {\bibinfo {title} {Glassy dynamics in a
  confined monatomic fluid},\ }\href
  {https://doi.org/10.1103/PhysRevE.86.011504} {\bibfield  {journal} {\bibinfo
  {journal} {Phys. Rev. E}\ }\textbf {\bibinfo {volume} {86}},\ \bibinfo
  {pages} {011504} (\bibinfo {year} {2012})}\BibitemShut {NoStop}%
\bibitem [{\citenamefont {Ingebrigtsen}\ \emph {et~al.}(2013)\citenamefont
  {Ingebrigtsen}, \citenamefont {Errington}, \citenamefont {Truskett},\ and\
  \citenamefont {Dyre}}]{sim:Ingebrigtsen2013}%
  \BibitemOpen
  \bibfield  {author} {\bibinfo {author} {\bibfnamefont {T.~S.}\ \bibnamefont
  {Ingebrigtsen}}, \bibinfo {author} {\bibfnamefont {J.~R.}\ \bibnamefont
  {Errington}}, \bibinfo {author} {\bibfnamefont {T.~M.}\ \bibnamefont
  {Truskett}},\ and\ \bibinfo {author} {\bibfnamefont {J.~C.}\ \bibnamefont
  {Dyre}},\ }\bibfield  {title} {\bibinfo {title} {Predicting how
  nanoconfinement changes the relaxation time of a supercooled liquid},\ }\href
  {https://doi.org/10.1103/PhysRevLett.111.235901} {\bibfield  {journal}
  {\bibinfo  {journal} {Phys. Rev. Lett.}\ }\textbf {\bibinfo {volume} {111}},\
  \bibinfo {pages} {235901} (\bibinfo {year} {2013})}\BibitemShut {NoStop}%
\bibitem [{\citenamefont {Saw}\ and\ \citenamefont
  {Dasgupta}(2016)}]{sim:Saw2016}%
  \BibitemOpen
  \bibfield  {author} {\bibinfo {author} {\bibfnamefont {S.}~\bibnamefont
  {Saw}}\ and\ \bibinfo {author} {\bibfnamefont {C.}~\bibnamefont {Dasgupta}},\
  }\bibfield  {title} {\bibinfo {title} {Role of density modulation in the
  spatially resolved dynamics of strongly confined liquids},\ }\href
  {https://doi.org/10.1063/1.4959942} {\bibfield  {journal} {\bibinfo
  {journal} {The Journal of Chemical Physics}\ }\textbf {\bibinfo {volume}
  {145}},\ \bibinfo {pages} {054707} (\bibinfo {year} {2016})}\BibitemShut
  {NoStop}%
\bibitem [{\citenamefont {Scheidler}\ \emph {et~al.}(2004)\citenamefont
  {Scheidler}, \citenamefont {Kob},\ and\ \citenamefont
  {Binder}}]{sim:Scheidler2004}%
  \BibitemOpen
  \bibfield  {author} {\bibinfo {author} {\bibfnamefont {P.}~\bibnamefont
  {Scheidler}}, \bibinfo {author} {\bibfnamefont {W.}~\bibnamefont {Kob}},\
  and\ \bibinfo {author} {\bibfnamefont {K.}~\bibnamefont {Binder}},\
  }\bibfield  {title} {\bibinfo {title} {The {Relaxation} {Dynamics} of a
  {Supercooled} {Liquid} {Confined} by {Rough} {Walls}},\ }\href
  {https://doi.org/10.1021/jp036593s} {\bibfield  {journal} {\bibinfo
  {journal} {J. Phys. Chem. B}\ }\textbf {\bibinfo {volume} {108}},\ \bibinfo
  {pages} {6673} (\bibinfo {year} {2004})}\BibitemShut {NoStop}%
\bibitem [{\citenamefont {Spannuth}\ and\ \citenamefont
  {Conrad}(2012)}]{PhysRevLett.109.028301}%
  \BibitemOpen
  \bibfield  {author} {\bibinfo {author} {\bibfnamefont {M.}~\bibnamefont
  {Spannuth}}\ and\ \bibinfo {author} {\bibfnamefont {J.~C.}\ \bibnamefont
  {Conrad}},\ }\bibfield  {title} {\bibinfo {title} {Confinement-induced
  solidification of colloid-polymer depletion mixtures},\ }\href
  {https://doi.org/10.1103/PhysRevLett.109.028301} {\bibfield  {journal}
  {\bibinfo  {journal} {Phys. Rev. Lett.}\ }\textbf {\bibinfo {volume} {109}},\
  \bibinfo {pages} {028301} (\bibinfo {year} {2012})}\BibitemShut {NoStop}%
\bibitem [{\citenamefont {Fasolo}\ and\ \citenamefont
  {Sollich}(2003)}]{polycrist:Fasolo2003}%
  \BibitemOpen
  \bibfield  {author} {\bibinfo {author} {\bibfnamefont {M.}~\bibnamefont
  {Fasolo}}\ and\ \bibinfo {author} {\bibfnamefont {P.}~\bibnamefont
  {Sollich}},\ }\bibfield  {title} {\bibinfo {title} {Equilibrium phase
  behavior of polydisperse hard spheres},\ }\href
  {https://doi.org/10.1103/PhysRevLett.91.068301} {\bibfield  {journal}
  {\bibinfo  {journal} {Phys. Rev. Lett.}\ }\textbf {\bibinfo {volume} {91}},\
  \bibinfo {pages} {068301} (\bibinfo {year} {2003})}\BibitemShut {NoStop}%
\bibitem [{\citenamefont {Zaccarelli}\ \emph {et~al.}(2009)\citenamefont
  {Zaccarelli}, \citenamefont {Valeriani}, \citenamefont {Sanz}, \citenamefont
  {Poon}, \citenamefont {Cates},\ and\ \citenamefont
  {Pusey}}]{polycrist:Zaccarelli2009}%
  \BibitemOpen
  \bibfield  {author} {\bibinfo {author} {\bibfnamefont {E.}~\bibnamefont
  {Zaccarelli}}, \bibinfo {author} {\bibfnamefont {C.}~\bibnamefont
  {Valeriani}}, \bibinfo {author} {\bibfnamefont {E.}~\bibnamefont {Sanz}},
  \bibinfo {author} {\bibfnamefont {W.~C.~K.}\ \bibnamefont {Poon}}, \bibinfo
  {author} {\bibfnamefont {M.~E.}\ \bibnamefont {Cates}},\ and\ \bibinfo
  {author} {\bibfnamefont {P.~N.}\ \bibnamefont {Pusey}},\ }\bibfield  {title}
  {\bibinfo {title} {Crystallization of hard-sphere glasses},\ }\href
  {https://doi.org/10.1103/PhysRevLett.103.135704} {\bibfield  {journal}
  {\bibinfo  {journal} {Phys. Rev. Lett.}\ }\textbf {\bibinfo {volume} {103}},\
  \bibinfo {pages} {135704} (\bibinfo {year} {2009})}\BibitemShut {NoStop}%
\bibitem [{\citenamefont {Campo}\ and\ \citenamefont
  {Speck}(2016)}]{polycrist:Campo2016}%
  \BibitemOpen
  \bibfield  {author} {\bibinfo {author} {\bibfnamefont {M.}~\bibnamefont
  {Campo}}\ and\ \bibinfo {author} {\bibfnamefont {T.}~\bibnamefont {Speck}},\
  }\bibfield  {title} {\bibinfo {title} {Polydisperse hard spheres:
  crystallization kinetics in small systems and role of local structure},\
  }\href {https://doi.org/10.1088/1742-5468/2016/8/084007} {\bibfield
  {journal} {\bibinfo  {journal} {Journal of Statistical Mechanics: Theory and
  Experiment}\ }\textbf {\bibinfo {volume} {2016}},\ \bibinfo {pages} {084007}
  (\bibinfo {year} {2016})}\BibitemShut {NoStop}%
\bibitem [{\citenamefont {Bommineni}\ \emph {et~al.}(2019)\citenamefont
  {Bommineni}, \citenamefont {Varela-Rosales}, \citenamefont {Klement},\ and\
  \citenamefont {Engel}}]{polycrist:Bommineni2019}%
  \BibitemOpen
  \bibfield  {author} {\bibinfo {author} {\bibfnamefont {P.~K.}\ \bibnamefont
  {Bommineni}}, \bibinfo {author} {\bibfnamefont {N.~R.}\ \bibnamefont
  {Varela-Rosales}}, \bibinfo {author} {\bibfnamefont {M.}~\bibnamefont
  {Klement}},\ and\ \bibinfo {author} {\bibfnamefont {M.}~\bibnamefont
  {Engel}},\ }\bibfield  {title} {\bibinfo {title} {Complex crystals from
  size-disperse spheres},\ }\href
  {https://doi.org/10.1103/PhysRevLett.122.128005} {\bibfield  {journal}
  {\bibinfo  {journal} {Phys. Rev. Lett.}\ }\textbf {\bibinfo {volume} {122}},\
  \bibinfo {pages} {128005} (\bibinfo {year} {2019})}\BibitemShut {NoStop}%
\bibitem [{\citenamefont {Hitchcock}\ and\ \citenamefont
  {Hall}(1999)}]{kacrist:Hitchcock1999}%
  \BibitemOpen
  \bibfield  {author} {\bibinfo {author} {\bibfnamefont {M.~R.}\ \bibnamefont
  {Hitchcock}}\ and\ \bibinfo {author} {\bibfnamefont {C.~K.}\ \bibnamefont
  {Hall}},\ }\bibfield  {title} {\bibinfo {title} {Solid–liquid phase
  equilibrium for binary lennard-jones mixtures},\ }\href
  {https://doi.org/10.1063/1.479084} {\bibfield  {journal} {\bibinfo  {journal}
  {The Journal of Chemical Physics}\ }\textbf {\bibinfo {volume} {110}},\
  \bibinfo {pages} {11433} (\bibinfo {year} {1999})}\BibitemShut {NoStop}%
\bibitem [{\citenamefont {Pedersen}\ \emph {et~al.}(2018)\citenamefont
  {Pedersen}, \citenamefont {Schr\o{}der},\ and\ \citenamefont
  {Dyre}}]{kacrist:Pedersen2018}%
  \BibitemOpen
  \bibfield  {author} {\bibinfo {author} {\bibfnamefont {U.~R.}\ \bibnamefont
  {Pedersen}}, \bibinfo {author} {\bibfnamefont {T.~B.}\ \bibnamefont
  {Schr\o{}der}},\ and\ \bibinfo {author} {\bibfnamefont {J.~C.}\ \bibnamefont
  {Dyre}},\ }\bibfield  {title} {\bibinfo {title} {Phase diagram of
  kob-andersen-type binary lennard-jones mixtures},\ }\href
  {https://doi.org/10.1103/PhysRevLett.120.165501} {\bibfield  {journal}
  {\bibinfo  {journal} {Phys. Rev. Lett.}\ }\textbf {\bibinfo {volume} {120}},\
  \bibinfo {pages} {165501} (\bibinfo {year} {2018})}\BibitemShut {NoStop}%
\bibitem [{\citenamefont {Sandomirski}\ \emph {et~al.}(2011)\citenamefont
  {Sandomirski}, \citenamefont {Allahyarov}, \citenamefont {Löwen},\ and\
  \citenamefont {Egelhaaf}}]{hetcry:Sandomirski2011}%
  \BibitemOpen
  \bibfield  {author} {\bibinfo {author} {\bibfnamefont {K.}~\bibnamefont
  {Sandomirski}}, \bibinfo {author} {\bibfnamefont {E.}~\bibnamefont
  {Allahyarov}}, \bibinfo {author} {\bibfnamefont {H.}~\bibnamefont {Löwen}},\
  and\ \bibinfo {author} {\bibfnamefont {S.~U.}\ \bibnamefont {Egelhaaf}},\
  }\bibfield  {title} {\bibinfo {title} {Heterogeneous crystallization of
  hard-sphere colloids near a wall},\ }\href
  {https://doi.org/10.1039/C1SM05346A} {\bibfield  {journal} {\bibinfo
  {journal} {Soft Matter}\ }\textbf {\bibinfo {volume} {7}},\ \bibinfo {pages}
  {8050} (\bibinfo {year} {2011})}\BibitemShut {NoStop}%
\bibitem [{\citenamefont {Espinosa}\ \emph {et~al.}(2019)\citenamefont
  {Espinosa}, \citenamefont {Vega}, \citenamefont {Valeriani}, \citenamefont
  {Frenkel},\ and\ \citenamefont {Sanz}}]{hetcry:Espinosa2019}%
  \BibitemOpen
  \bibfield  {author} {\bibinfo {author} {\bibfnamefont {J.~R.}\ \bibnamefont
  {Espinosa}}, \bibinfo {author} {\bibfnamefont {C.}~\bibnamefont {Vega}},
  \bibinfo {author} {\bibfnamefont {C.}~\bibnamefont {Valeriani}}, \bibinfo
  {author} {\bibfnamefont {D.}~\bibnamefont {Frenkel}},\ and\ \bibinfo {author}
  {\bibfnamefont {E.}~\bibnamefont {Sanz}},\ }\bibfield  {title} {\bibinfo
  {title} {Heterogeneous versus homogeneous crystal nucleation of hard
  spheres},\ }\href {https://doi.org/10.1039/C9SM01142K} {\bibfield  {journal}
  {\bibinfo  {journal} {Soft Matter}\ }\textbf {\bibinfo {volume} {15}},\
  \bibinfo {pages} {9625} (\bibinfo {year} {2019})}\BibitemShut {NoStop}%
\bibitem [{\citenamefont {Alder}\ and\ \citenamefont
  {Wainwright}(1957)}]{edmd:alder1957}%
  \BibitemOpen
  \bibfield  {author} {\bibinfo {author} {\bibfnamefont {B.~J.}\ \bibnamefont
  {Alder}}\ and\ \bibinfo {author} {\bibfnamefont {T.~E.}\ \bibnamefont
  {Wainwright}},\ }\bibfield  {title} {\bibinfo {title} {Phase transition for a
  hard sphere system},\ }\href {https://doi.org/10.1063/1.1743957} {\bibfield
  {journal} {\bibinfo  {journal} {The Journal of Chemical Physics}\ }\textbf
  {\bibinfo {volume} {27}},\ \bibinfo {pages} {1208} (\bibinfo {year}
  {1957})}\BibitemShut {NoStop}%
\bibitem [{\citenamefont {Rapaport}(1980)}]{edmd:RAPAPORT1980}%
  \BibitemOpen
  \bibfield  {author} {\bibinfo {author} {\bibfnamefont {D.}~\bibnamefont
  {Rapaport}},\ }\bibfield  {title} {\bibinfo {title} {The event scheduling
  problem in molecular dynamic simulation},\ }\href
  {https://doi.org/https://doi.org/10.1016/0021-9991(80)90104-7} {\bibfield
  {journal} {\bibinfo  {journal} {Journal of Computational Physics}\ }\textbf
  {\bibinfo {volume} {34}},\ \bibinfo {pages} {184 } (\bibinfo {year}
  {1980})}\BibitemShut {NoStop}%
\bibitem [{\citenamefont {Bannerman}\ \emph {et~al.}(2011)\citenamefont
  {Bannerman}, \citenamefont {Sargant},\ and\ \citenamefont
  {Lue}}]{edmd:bannerman2011}%
  \BibitemOpen
  \bibfield  {author} {\bibinfo {author} {\bibfnamefont {M.~N.}\ \bibnamefont
  {Bannerman}}, \bibinfo {author} {\bibfnamefont {R.}~\bibnamefont {Sargant}},\
  and\ \bibinfo {author} {\bibfnamefont {L.}~\bibnamefont {Lue}},\ }\bibfield
  {title} {\bibinfo {title} {Dynamo: a free $\mathcal{O}(n)$ general
  event-driven molecular dynamics simulator},\ }\href
  {https://doi.org/10.1002/jcc.21915} {\bibfield  {journal} {\bibinfo
  {journal} {Journal of Computational Chemistry}\ }\textbf {\bibinfo {volume}
  {32}},\ \bibinfo {pages} {3329} (\bibinfo {year} {2011})}\BibitemShut
  {NoStop}%
\bibitem [{\citenamefont {Gazzillo}\ and\ \citenamefont
  {Pastore}(1989)}]{swap:GAZZILLO1989}%
  \BibitemOpen
  \bibfield  {author} {\bibinfo {author} {\bibfnamefont {D.}~\bibnamefont
  {Gazzillo}}\ and\ \bibinfo {author} {\bibfnamefont {G.}~\bibnamefont
  {Pastore}},\ }\bibfield  {title} {\bibinfo {title} {Equation of state for
  symmetric non-additive hard-sphere fluids: An approximate analytic expression
  and new monte carlo results},\ }\href
  {https://doi.org/https://doi.org/10.1016/0009-2614(89)87505-0} {\bibfield
  {journal} {\bibinfo  {journal} {Chemical Physics Letters}\ }\textbf {\bibinfo
  {volume} {159}},\ \bibinfo {pages} {388 } (\bibinfo {year}
  {1989})}\BibitemShut {NoStop}%
\bibitem [{\citenamefont {Ninarello}\ \emph {et~al.}(2017)\citenamefont
  {Ninarello}, \citenamefont {Berthier},\ and\ \citenamefont
  {Coslovich}}]{swap:ninarello2017}%
  \BibitemOpen
  \bibfield  {author} {\bibinfo {author} {\bibfnamefont {A.}~\bibnamefont
  {Ninarello}}, \bibinfo {author} {\bibfnamefont {L.}~\bibnamefont
  {Berthier}},\ and\ \bibinfo {author} {\bibfnamefont {D.}~\bibnamefont
  {Coslovich}},\ }\bibfield  {title} {\bibinfo {title} {Models and algorithms
  for the next generation of glass transition studies},\ }\href
  {https://doi.org/10.1103/PhysRevX.7.021039} {\bibfield  {journal} {\bibinfo
  {journal} {Phys. Rev. X}\ }\textbf {\bibinfo {volume} {7}},\ \bibinfo {pages}
  {021039} (\bibinfo {year} {2017})}\BibitemShut {NoStop}%
\bibitem [{\citenamefont {Berthier}\ \emph {et~al.}(2019)\citenamefont
  {Berthier}, \citenamefont {Flenner}, \citenamefont {Fullerton}, \citenamefont
  {Scalliet},\ and\ \citenamefont {Singh}}]{swap:Berthier2019}%
  \BibitemOpen
  \bibfield  {author} {\bibinfo {author} {\bibfnamefont {L.}~\bibnamefont
  {Berthier}}, \bibinfo {author} {\bibfnamefont {E.}~\bibnamefont {Flenner}},
  \bibinfo {author} {\bibfnamefont {C.~J.}\ \bibnamefont {Fullerton}}, \bibinfo
  {author} {\bibfnamefont {C.}~\bibnamefont {Scalliet}},\ and\ \bibinfo
  {author} {\bibfnamefont {M.}~\bibnamefont {Singh}},\ }\bibfield  {title}
  {\bibinfo {title} {Efficient swap algorithms for molecular dynamics
  simulations of equilibrium supercooled liquids},\ }\href
  {https://doi.org/10.1088/1742-5468/ab1910} {\bibfield  {journal} {\bibinfo
  {journal} {Journal of Statistical Mechanics: Theory and Experiment}\ }\textbf
  {\bibinfo {volume} {2019}},\ \bibinfo {pages} {064004} (\bibinfo {year}
  {2019})}\BibitemShut {NoStop}%
\bibitem [{\citenamefont {Nelson}\ and\ \citenamefont
  {Halperin}(1979)}]{bo:Nelson1979}%
  \BibitemOpen
  \bibfield  {author} {\bibinfo {author} {\bibfnamefont {D.~R.}\ \bibnamefont
  {Nelson}}\ and\ \bibinfo {author} {\bibfnamefont {B.~I.}\ \bibnamefont
  {Halperin}},\ }\bibfield  {title} {\bibinfo {title} {Dislocation-mediated
  melting in two dimensions},\ }\href
  {https://doi.org/10.1103/PhysRevB.19.2457} {\bibfield  {journal} {\bibinfo
  {journal} {Phys. Rev. B}\ }\textbf {\bibinfo {volume} {19}},\ \bibinfo
  {pages} {2457} (\bibinfo {year} {1979})}\BibitemShut {NoStop}%
\bibitem [{\citenamefont {Chakrabarti}\ and\ \citenamefont
  {L\"owen}(1998)}]{bo:Chakrabarti1998}%
  \BibitemOpen
  \bibfield  {author} {\bibinfo {author} {\bibfnamefont {J.}~\bibnamefont
  {Chakrabarti}}\ and\ \bibinfo {author} {\bibfnamefont {H.}~\bibnamefont
  {L\"owen}},\ }\bibfield  {title} {\bibinfo {title} {Effect of confinement on
  charge-stabilized colloidal suspensions between two charged plates},\ }\href
  {https://doi.org/10.1103/PhysRevE.58.3400} {\bibfield  {journal} {\bibinfo
  {journal} {Phys. Rev. E}\ }\textbf {\bibinfo {volume} {58}},\ \bibinfo
  {pages} {3400} (\bibinfo {year} {1998})}\BibitemShut {NoStop}%
\bibitem [{\citenamefont {Pusey}\ \emph {et~al.}(2009)\citenamefont {Pusey},
  \citenamefont {Zaccarelli}, \citenamefont {Valeriani}, \citenamefont {Sanz},
  \citenamefont {Poon},\ and\ \citenamefont {Cates}}]{bo:Pusey2009}%
  \BibitemOpen
  \bibfield  {author} {\bibinfo {author} {\bibfnamefont {P.~N.}\ \bibnamefont
  {Pusey}}, \bibinfo {author} {\bibfnamefont {E.}~\bibnamefont {Zaccarelli}},
  \bibinfo {author} {\bibfnamefont {C.}~\bibnamefont {Valeriani}}, \bibinfo
  {author} {\bibfnamefont {E.}~\bibnamefont {Sanz}}, \bibinfo {author}
  {\bibfnamefont {W.~C.~K.}\ \bibnamefont {Poon}},\ and\ \bibinfo {author}
  {\bibfnamefont {M.~E.}\ \bibnamefont {Cates}},\ }\bibfield  {title} {\bibinfo
  {title} {Hard spheres: crystallization and glass formation},\ }\href
  {https://doi.org/10.1098/rsta.2009.0181} {\bibfield  {journal} {\bibinfo
  {journal} {Philosophical Transactions of the Royal Society A: Mathematical,
  Physical and Engineering Sciences}\ }\textbf {\bibinfo {volume} {367}},\
  \bibinfo {pages} {4993} (\bibinfo {year} {2009})}\BibitemShut {NoStop}%
\bibitem [{\citenamefont {Rosenfeld}(1989)}]{fmt:Rosenfeld1989}%
  \BibitemOpen
  \bibfield  {author} {\bibinfo {author} {\bibfnamefont {Y.}~\bibnamefont
  {Rosenfeld}},\ }\bibfield  {title} {\bibinfo {title} {Free-energy model for
  the inhomogeneous hard-sphere fluid mixture and density-functional theory of
  freezing},\ }\href {https://doi.org/10.1103/PhysRevLett.63.980} {\bibfield
  {journal} {\bibinfo  {journal} {Phys. Rev. Lett.}\ }\textbf {\bibinfo
  {volume} {63}},\ \bibinfo {pages} {980} (\bibinfo {year} {1989})}\BibitemShut
  {NoStop}%
\bibitem [{\citenamefont {Roth}(2010)}]{fmt:Roth2010}%
  \BibitemOpen
  \bibfield  {author} {\bibinfo {author} {\bibfnamefont {R.}~\bibnamefont
  {Roth}},\ }\bibfield  {title} {\bibinfo {title} {Fundamental measure theory
  for hard-sphere mixtures: a review},\ }\href
  {https://doi.org/10.1088/0953-8984/22/6/063102} {\bibfield  {journal}
  {\bibinfo  {journal} {Journal of Physics: Condensed Matter}\ }\textbf
  {\bibinfo {volume} {22}},\ \bibinfo {pages} {063102} (\bibinfo {year}
  {2010})}\BibitemShut {NoStop}%
\bibitem [{\citenamefont {Heyes}\ and\ \citenamefont
  {Santos}(2018)}]{fmt:chemical2018}%
  \BibitemOpen
  \bibfield  {author} {\bibinfo {author} {\bibfnamefont {D.~M.}\ \bibnamefont
  {Heyes}}\ and\ \bibinfo {author} {\bibfnamefont {A.}~\bibnamefont {Santos}},\
  }\bibfield  {title} {\bibinfo {title} {Chemical potential of a test hard
  sphere of variable size in hard-sphere fluid mixtures},\ }\href
  {https://doi.org/10.1063/1.5037856} {\bibfield  {journal} {\bibinfo
  {journal} {The Journal of Chemical Physics}\ }\textbf {\bibinfo {volume}
  {148}},\ \bibinfo {pages} {214503} (\bibinfo {year} {2018})}\BibitemShut
  {NoStop}%
\bibitem [{\citenamefont {Kirkwood}(1950)}]{cell:kirkwood1950}%
  \BibitemOpen
  \bibfield  {author} {\bibinfo {author} {\bibfnamefont {J.~G.}\ \bibnamefont
  {Kirkwood}},\ }\bibfield  {title} {\bibinfo {title} {Critique of the free
  volume theory of the liquid state},\ }\href
  {https://doi.org/10.1063/1.1747635} {\bibfield  {journal} {\bibinfo
  {journal} {The Journal of Chemical Physics}\ }\textbf {\bibinfo {volume}
  {18}},\ \bibinfo {pages} {380} (\bibinfo {year} {1950})}\BibitemShut
  {NoStop}%
\bibitem [{\citenamefont {Barker}(1963)}]{barker1963lattice}%
  \BibitemOpen
  \bibfield  {author} {\bibinfo {author} {\bibfnamefont {J.~A.}\ \bibnamefont
  {Barker}},\ }\href@noop {} {\emph {\bibinfo {title} {Lattice theories of the
  liquid state}}},\ Vol.~\bibinfo {volume} {1}\ (\bibinfo  {publisher}
  {Pergamon Press},\ \bibinfo {year} {1963})\BibitemShut {NoStop}%
\bibitem [{\citenamefont {Bonissent}\ \emph {et~al.}(1984)\citenamefont
  {Bonissent}, \citenamefont {Pieranski},\ and\ \citenamefont
  {Pieranski}}]{cell:Bonissent1984}%
  \BibitemOpen
  \bibfield  {author} {\bibinfo {author} {\bibfnamefont {A.}~\bibnamefont
  {Bonissent}}, \bibinfo {author} {\bibfnamefont {P.}~\bibnamefont
  {Pieranski}},\ and\ \bibinfo {author} {\bibfnamefont {P.}~\bibnamefont
  {Pieranski}},\ }\bibfield  {title} {\bibinfo {title} {Solid-solid phase
  transitions in a low-dimensionality system},\ }\href
  {https://doi.org/10.1080/01418618408244211} {\bibfield  {journal} {\bibinfo
  {journal} {Philosophical Magazine A}\ }\textbf {\bibinfo {volume} {50}},\
  \bibinfo {pages} {57} (\bibinfo {year} {1984})}\BibitemShut {NoStop}%
\bibitem [{\citenamefont {Salacuse}\ and\ \citenamefont
  {Stell}(1982)}]{mix:Salacuse1982}%
  \BibitemOpen
  \bibfield  {author} {\bibinfo {author} {\bibfnamefont {J.~J.}\ \bibnamefont
  {Salacuse}}\ and\ \bibinfo {author} {\bibfnamefont {G.}~\bibnamefont
  {Stell}},\ }\bibfield  {title} {\bibinfo {title} {Polydisperse systems:
  Statistical thermodynamics, with applications to several models including
  hard and permeable spheres},\ }\href {https://doi.org/10.1063/1.444274}
  {\bibfield  {journal} {\bibinfo  {journal} {The Journal of Chemical Physics}\
  }\textbf {\bibinfo {volume} {77}},\ \bibinfo {pages} {3714} (\bibinfo {year}
  {1982})}\BibitemShut {NoStop}%
\bibitem [{\citenamefont {Sear}(1998)}]{mix:Sear_1998}%
  \BibitemOpen
  \bibfield  {author} {\bibinfo {author} {\bibfnamefont {R.~P.}\ \bibnamefont
  {Sear}},\ }\bibfield  {title} {\bibinfo {title} {Phase separation and
  crystallisation of polydisperse hard spheres},\ }\href
  {https://doi.org/10.1209/epl/i1998-00500-3} {\bibfield  {journal} {\bibinfo
  {journal} {Europhysics Letters ({EPL})}\ }\textbf {\bibinfo {volume} {44}},\
  \bibinfo {pages} {531} (\bibinfo {year} {1998})}\BibitemShut {NoStop}%
\bibitem [{\citenamefont {Sollich}\ \emph {et~al.}(2001)\citenamefont
  {Sollich}, \citenamefont {Warren},\ and\ \citenamefont
  {Cates}}]{theo:Sollich2007}%
  \BibitemOpen
  \bibfield  {author} {\bibinfo {author} {\bibfnamefont {P.}~\bibnamefont
  {Sollich}}, \bibinfo {author} {\bibfnamefont {P.~B.}\ \bibnamefont
  {Warren}},\ and\ \bibinfo {author} {\bibfnamefont {M.~E.}\ \bibnamefont
  {Cates}},\ }\bibinfo {title} {Moment free energies for polydisperse
  systems},\ in\ \href {https://doi.org/10.1002/9780470141762.ch4} {\emph
  {\bibinfo {booktitle} {Advances in Chemical Physics}}}\ (\bibinfo
  {publisher} {John Wiley and Sons, Ltd},\ \bibinfo {year} {2001})\ pp.\
  \bibinfo {pages} {265--336}\BibitemShut {NoStop}%
\bibitem [{\citenamefont {Pieranski}\ \emph {et~al.}(1983)\citenamefont
  {Pieranski}, \citenamefont {Strzelecki},\ and\ \citenamefont
  {Pansu}}]{exp:Pieranski1983}%
  \BibitemOpen
  \bibfield  {author} {\bibinfo {author} {\bibfnamefont {P.}~\bibnamefont
  {Pieranski}}, \bibinfo {author} {\bibfnamefont {L.}~\bibnamefont
  {Strzelecki}},\ and\ \bibinfo {author} {\bibfnamefont {B.}~\bibnamefont
  {Pansu}},\ }\bibfield  {title} {\bibinfo {title} {Thin colloidal crystals},\
  }\href {https://doi.org/10.1103/PhysRevLett.50.900} {\bibfield  {journal}
  {\bibinfo  {journal} {Phys. Rev. Lett.}\ }\textbf {\bibinfo {volume} {50}},\
  \bibinfo {pages} {900} (\bibinfo {year} {1983})}\BibitemShut {NoStop}%
\bibitem [{\citenamefont {Van~Winkle}\ and\ \citenamefont
  {Murray}(1986)}]{exp:Winkle1986}%
  \BibitemOpen
  \bibfield  {author} {\bibinfo {author} {\bibfnamefont {D.~H.}\ \bibnamefont
  {Van~Winkle}}\ and\ \bibinfo {author} {\bibfnamefont {C.~A.}\ \bibnamefont
  {Murray}},\ }\bibfield  {title} {\bibinfo {title} {Layering transitions in
  colloidal crystals as observed by diffraction and direct-lattice imaging},\
  }\href {https://doi.org/10.1103/PhysRevA.34.562} {\bibfield  {journal}
  {\bibinfo  {journal} {Phys. Rev. A}\ }\textbf {\bibinfo {volume} {34}},\
  \bibinfo {pages} {562} (\bibinfo {year} {1986})}\BibitemShut {NoStop}%
\bibitem [{\citenamefont {Bubeck}\ \emph {et~al.}(1999)\citenamefont {Bubeck},
  \citenamefont {Bechinger}, \citenamefont {Neser},\ and\ \citenamefont
  {Leiderer}}]{reent:Bechinger1999}%
  \BibitemOpen
  \bibfield  {author} {\bibinfo {author} {\bibfnamefont {R.}~\bibnamefont
  {Bubeck}}, \bibinfo {author} {\bibfnamefont {C.}~\bibnamefont {Bechinger}},
  \bibinfo {author} {\bibfnamefont {S.}~\bibnamefont {Neser}},\ and\ \bibinfo
  {author} {\bibfnamefont {P.}~\bibnamefont {Leiderer}},\ }\bibfield  {title}
  {\bibinfo {title} {Melting and reentrant freezing of two-dimensional
  colloidal crystals in confined geometry},\ }\href
  {https://doi.org/10.1103/PhysRevLett.82.3364} {\bibfield  {journal} {\bibinfo
   {journal} {Phys. Rev. Lett.}\ }\textbf {\bibinfo {volume} {82}},\ \bibinfo
  {pages} {3364} (\bibinfo {year} {1999})}\BibitemShut {NoStop}%
\bibitem [{\citenamefont {Ohara}\ \emph {et~al.}(1995)\citenamefont {Ohara},
  \citenamefont {Leff}, \citenamefont {Heath},\ and\ \citenamefont
  {Gelbart}}]{exp:opal1995}%
  \BibitemOpen
  \bibfield  {author} {\bibinfo {author} {\bibfnamefont {P.~C.}\ \bibnamefont
  {Ohara}}, \bibinfo {author} {\bibfnamefont {D.~V.}\ \bibnamefont {Leff}},
  \bibinfo {author} {\bibfnamefont {J.~R.}\ \bibnamefont {Heath}},\ and\
  \bibinfo {author} {\bibfnamefont {W.~M.}\ \bibnamefont {Gelbart}},\
  }\bibfield  {title} {\bibinfo {title} {Crystallization of opals from
  polydisperse nanoparticles},\ }\href
  {https://doi.org/10.1103/PhysRevLett.75.3466} {\bibfield  {journal} {\bibinfo
   {journal} {Phys. Rev. Lett.}\ }\textbf {\bibinfo {volume} {75}},\ \bibinfo
  {pages} {3466} (\bibinfo {year} {1995})}\BibitemShut {NoStop}%
\bibitem [{\citenamefont {Klaver}\ and\ \citenamefont
  {Schroën}(2015)}]{shear2015}%
  \BibitemOpen
  \bibfield  {author} {\bibinfo {author} {\bibfnamefont {R.}~\bibnamefont
  {Klaver}}\ and\ \bibinfo {author} {\bibfnamefont {C.}~\bibnamefont
  {Schroën}},\ }\bibfield  {title} {\bibinfo {title} {A review of
  shear-induced particle migration for enhanced filtration and fractionation},\
  }in\ \href
  {https://doi.org/https://doi.org/10.1016/B978-1-78242-284-6.00008-8} {\emph
  {\bibinfo {booktitle} {Modeling Food Processing Operations}}},\ \bibinfo
  {series and number} {Woodhead Publishing Series in Food Science, Technology
  and Nutrition}\ (\bibinfo  {publisher} {Woodhead Publishing},\ \bibinfo
  {year} {2015})\ pp.\ \bibinfo {pages} {211 -- 233}\BibitemShut {NoStop}%
\bibitem [{\citenamefont {Wu}\ \emph {et~al.}(2009)\citenamefont {Wu},
  \citenamefont {Derks}, \citenamefont {van Blaaderen},\ and\ \citenamefont
  {Imhof}}]{shear2009}%
  \BibitemOpen
  \bibfield  {author} {\bibinfo {author} {\bibfnamefont {Y.~L.}\ \bibnamefont
  {Wu}}, \bibinfo {author} {\bibfnamefont {D.}~\bibnamefont {Derks}}, \bibinfo
  {author} {\bibfnamefont {A.}~\bibnamefont {van Blaaderen}},\ and\ \bibinfo
  {author} {\bibfnamefont {A.}~\bibnamefont {Imhof}},\ }\bibfield  {title}
  {\bibinfo {title} {Melting and crystallization of colloidal hard-sphere
  suspensions under shear},\ }\href {https://doi.org/10.1073/pnas.0812519106}
  {\bibfield  {journal} {\bibinfo  {journal} {Proceedings of the National
  Academy of Sciences}\ }\textbf {\bibinfo {volume} {106}},\ \bibinfo {pages}
  {10564} (\bibinfo {year} {2009})},\ \Eprint
  {https://arxiv.org/abs/https://www.pnas.org/content/106/26/10564.full.pdf}
  {https://www.pnas.org/content/106/26/10564.full.pdf} \BibitemShut {NoStop}%
\bibitem [{\citenamefont {Iacopini}\ \emph {et~al.}(2009)\citenamefont
  {Iacopini}, \citenamefont {Palberg},\ and\ \citenamefont
  {Sch\"ope}}]{exp:Palberg2009}%
  \BibitemOpen
  \bibfield  {author} {\bibinfo {author} {\bibfnamefont {S.}~\bibnamefont
  {Iacopini}}, \bibinfo {author} {\bibfnamefont {T.}~\bibnamefont {Palberg}},\
  and\ \bibinfo {author} {\bibfnamefont {H.~J.}\ \bibnamefont {Sch\"ope}},\
  }\bibfield  {title} {\bibinfo {title} {Ripening-dominated crystallization in
  polydisperse hard-sphere-like colloids},\ }\href
  {https://doi.org/10.1103/PhysRevE.79.010601} {\bibfield  {journal} {\bibinfo
  {journal} {Phys. Rev. E}\ }\textbf {\bibinfo {volume} {79}},\ \bibinfo
  {pages} {010601(R)} (\bibinfo {year} {2009})}\BibitemShut {NoStop}%
\bibitem [{\citenamefont {Cabane}\ \emph {et~al.}(2016)\citenamefont {Cabane},
  \citenamefont {Li}, \citenamefont {Artzner}, \citenamefont {Botet},
  \citenamefont {Labbez}, \citenamefont {Bareigts}, \citenamefont {Sztucki},\
  and\ \citenamefont {Goehring}}]{exp:chargedCol2016}%
  \BibitemOpen
  \bibfield  {author} {\bibinfo {author} {\bibfnamefont {B.}~\bibnamefont
  {Cabane}}, \bibinfo {author} {\bibfnamefont {J.}~\bibnamefont {Li}}, \bibinfo
  {author} {\bibfnamefont {F.}~\bibnamefont {Artzner}}, \bibinfo {author}
  {\bibfnamefont {R.}~\bibnamefont {Botet}}, \bibinfo {author} {\bibfnamefont
  {C.}~\bibnamefont {Labbez}}, \bibinfo {author} {\bibfnamefont
  {G.}~\bibnamefont {Bareigts}}, \bibinfo {author} {\bibfnamefont
  {M.}~\bibnamefont {Sztucki}},\ and\ \bibinfo {author} {\bibfnamefont
  {L.}~\bibnamefont {Goehring}},\ }\bibfield  {title} {\bibinfo {title} {Hiding
  in plain view: Colloidal self-assembly from polydisperse populations},\
  }\href {https://doi.org/10.1103/PhysRevLett.116.208001} {\bibfield  {journal}
  {\bibinfo  {journal} {Phys. Rev. Lett.}\ }\textbf {\bibinfo {volume} {116}},\
  \bibinfo {pages} {208001} (\bibinfo {year} {2016})}\BibitemShut {NoStop}%
\end{thebibliography}%
	
\end{document}